\pdfoutput=1

\documentclass[times,review]{elsarticle}

\usepackage{jcomp}
\usepackage{framed,multirow}

\usepackage[ruled,vlined]{algorithm2e}
\usepackage{amsmath}
\usepackage{amssymb}
\usepackage{latexsym}
\usepackage{flafter}
\usepackage{array}   
\newcolumntype{L}{>{$}l<{$}} 
\newcolumntype{C}{>{$}c<{$}} 

\usepackage[percent]{overpic}
\usepackage{url}
\usepackage[dvipsnames]{xcolor}
\definecolor{newcolor}{rgb}{.8,.349,.1}

\journal{Journal of Computational Physics}

\begin{document}

\newcommand{\ke}[1]{\textcolor{blue}{#1}}
\newcommand{\tim}[1]{\textcolor{red}{#1}}
\newcommand{\benedikt}[1]{\textcolor{ForestGreen}{#1}}

\def\vx{\mathbf{x}}
\def\vy{\mathbf{y}}
\def\vr{\mathsf{r}}
\def\vk{\mathsf{k}}
\def\vu{\mathsf{u}}
\def\vv{\mathsf{v}}
\def\va{\mathsf{a}}
\def\vb{\mathsf{b}}
\def\vc{\mathsf{c}}
\def\vn{{\mathbf{n}}}
\def\vp{\mathsf{p}}
\def\vP{\mathsf{P}}
\def\vq{\mathsf{q}}
\def\vw{\mathsf{w}}
\def\vg{\mathsf{g}}
\def\vf{\mathsf{f}}
\def\vphi{\mathsf{\phi}}
\def\vepsilon{\mathsf{\epsilon}}
\def\vs{\mathsf{s}}
\def\vS{\mathsf{S}}
\def\vI{\boldsymbol{I}}
\def\vA{\boldsymbol{A}}
\def\vGamma{{\mathsf{\Gamma}}}

\def\rmx{\mathrm{x}}
\def\rmy{\mathrm{y}}
\def\rmr{\mathrm{r}}
\def\rmk{\mathrm{k}}
\def\rmu{\mathrm{u}}
\def\rmv{\mathrm{v}}
\def\rma{\mathrm{a}}
\def\rmb{\mathrm{b}}
\def\rmc{\mathrm{c}}
\def\rmn{\mathrm{n}}

\def\vrmx{\mathbf{\mathrm{x}}}
\def\vrmy{\mathbf{\mathrm{y}}}
\def\vrmr{\mathbf{\mathrm{r}}}
\def\vrmk{\mathbf{\mathrm{k}}}
\def\vrmu{\mathbf{\mathrm{u}}}
\def\vrmv{\mathbf{\mathrm{v}}}
\def\vrma{\mathbf{\mathrm{a}}}
\def\vrmb{\mathbf{\mathrm{b}}}
\def\vrmc{\mathbf{\mathrm{c}}}
\def\vrmn{\mathbf{\mathrm{n}}}
\def\vrmI{\boldsymbol{\mathrm{I}}}
\def\vrmA{\boldsymbol{\mathrm{A}}}

\def\ven{\hat{\boldsymbol{n}}}
\def\hf{\hat{\boldsymbol{f}}}
\def\hG{\hat{\boldsymbol{G}}}
\def\vpsi{\boldsymbol{ \psi}}
\def\vomega{\boldsymbol{ \omega}}
\def\hvpsi{\hat{\mbox{\boldsymbol{ \psi}}}}
\def\hvo{\hat{{\mbox{\boldsymbol{ \omega}}}}}
\def\hvu{\hat{\vu}}
\def\cS{{\cal S}}
\def\cN{{\cal N}}
\def\cH{{\cal H}}
\def\cE{{\cal E}}
\def\cC{{\cal C}}
\def\cO{{\cal O}}
\def\cF{{\cal F}}
\def\cQ{{\cal Q}}
\def\cV{{\cal V}}

\def\rmL{{\mathsf{L}}}
\def\rmA{{\mathsf{A}}}
\def\rmH{{\mathsf{H}}}
\def\rmD{{\mathsf{D}}}
\def\rmG{{\mathsf{G}}}
\def\rmC{{\mathsf{C}}}
\def\rmN{{\mathsf{N}}}
\def\rmT{{\mathsf{T}}}

\def\R{\mathbb{R}}
\newcommand\Rey{\operatorname{Re}}

\newcommand{\intdiff}{\mathop{}\!\mathrm{d}}

\verso{K. Yu, B. Dorschner, T. Colonius}

\begin{frontmatter}

\title{Multi-resolution lattice Green's function method for incompressible flows}%

\author[1]{Ke Yu\corref{cor1}}
\cortext[cor1]{Corresponding author.}
\ead{
kyu2@caltech.edu
}

\author[1]{Benedikt Dorschner}
\ead{
bdorschn@ethz.ch
}

\author[1]{Tim Colonius}
\ead{
colonius@caltech.edu
}

\address[1]{
Department of Mechanical and Civil Engineering, California Institute of Technology,
Pasadena, CA 91125, USA
}

\received{Oct 2020}

\begin{abstract}
We propose a multi-resolution strategy that is compatible with the lattice Green’s function (LGF) technique for solving viscous, incompressible flows on unbounded domains.
The LGF method exploits the regularity of a finite-volume scheme on a formally unbounded Cartesian mesh to yield robust and computationally efficient solutions.  The original method is spatially adaptive, but challenging to integrate with embedded mesh refinement as the underlying LGF is only defined for a fixed resolution.  We present an ansatz for adaptive mesh refinement, where the solutions to the pressure Poisson equation are approximated using the LGF technique on a composite mesh constructed from a series of infinite lattices of differing resolution. To solve the incompressible Navier-Stokes equations, this is further combined with an integrating factor for the viscous terms and an appropriate Runge-Kutta scheme for the resulting differential-algebraic equations. The parallelized algorithm is verified through with numerical simulations of vortex rings, and the collision of vortex rings at high Reynolds number is simulated to demonstrate the reduction in computational cells achievable with both spatial and refinement adaptivity.
\end{abstract}


\end{frontmatter}

\section{Introduction}

Numerical simulations of high Reynolds numbers, incompressible flows on unbounded domains are challenging due to the wide range of physical scales and unbounded computational domain. The scale separation associated with the boundary layers and turbulence favors a flexible distribution of elements/cells with refinement in regions of high gradients.  For unstructured or structured body-fitted meshes, many techniques are available to achieve this clustering of elements, whereas for immersed boundary methods, the most natural way to do this is through static or adaptive mesh refinement (AMR) in  \cite{berger1984,berger_local_1989,macneice_paramesh_2000,nissen_stable_2015}.

The unbounded domain may be truncated with artificial inflow/outflow boundary conditions \cite{lackner_computation_1976,james_solution_1977, tsynkov1998numerical,colonius_modeling_2004,mccorquodale_local_2007}, or truncated based on the compact vorticity field together with a free-space Green's function that satisfies the exact far-field boundary condition.  The Green's function can be based on the discretized equations, i.e., the lattice Green's function (LGF) \cite{glasser_extended_1977, delves_green_2001, gillman_fast_2010,  gillman_fast_2014,liska2014parallel,liska2016fast,liska_fast_2017}, or the continuous ones, as is typically done in vortex methods (e.g.,  \cite{leonard_vortex_1980, chatelain_fourier_based_2010}). The former LGF has the advantage that, when combined with mimetic finite-difference/finite-volume methods, the resulting schemes are discretely conservative with provable stability bounds \cite{liska2016fast}.  Efficient, parallel solutions of the discrete convolution in three spatial dimensions can be achieved by adapting variants of the fast multipole method \cite{liska2014parallel}, which we refer to as the fast LGF (FLGF) method.

The FLGF based Navier-Stokes scheme can be made spatially adaptive by truncating the computation based on thresholding the vorticity or the Lamb vector, providing optimal spatial coverage of cells in important vortical flow regions \cite{liska2016fast}.  It can be fruitfully combined with the immersed boundary method to handle immersed surfaces on the regular lattice \cite{liska_fast_2017}, but an important disadvantage is that the LGF does not readily permit the static or adaptive local refinement required to efficiently simulate high Reynolds number flows.  While several multi-resolution schemes based on FMM and multigrid have been proposed for the Poisson equation that arises in incompressible flow (e.g. \cite{lashuk2009massivelyFMM, ying_kernel-independent_2004}), these methods are likewise not straightforward to combine with the LGF.

In our recent work \cite{dorschner2020fast}, we proposed a multi-resolution extension of the FLGF method (FLGF-AMR) that enables
block-structured mesh refinement while retaining the efficiency of the FLGF technique. In the present paper, we propose an ansatz for AMR that reinterprets and improves this algorithm, and we further extend the technique to solve the incompressible Navier-Stokes equations.  We consider the AMR grid as a restriction from an ambient composite grid that is constructed from a series of infinite lattices of differing resolution. Solutions to the Poisson equation are formally solved on every level of the composite mesh using the LGF, before being restricted back to the AMR gird.  We then construct commutative interpolation operators that obviate the need for explicitly computing most of the composite grid.  
In applying the scheme to the full Navier-Stokes equations,  we limit our attention to unbounded flows without immersed surfaces, but the algorithms we propose are compatible with the previous IBLGF method and will be combined in future work.

The paper is organized as follows.
In section~\ref{sec:uniform}, the FLGF based scheme for solving incompressible flows on unbounded uniform grids is briefly reviewed.
In section \ref{sec:AMR}, the concept of a composite grid is introduced. The previous FLGF-AMR method \cite{dorschner2020fast} is recast in this framework and an extended source correction is proposed. In section \ref{sec:FIF-AMR} we extend this framework to the LGF for an integrating factor to exactly advance the viscous terms when combined with a half-explicit Runge-Kutta scheme, and in \ref{sec:NS-LGF-AMR} we construct the remaining operators for the multi-resolution Navier-Stokes solver for incompressible external flows. Section \ref{sec:adaptivity} discusses how both spatial and refinement adaptivity can be achieved and section \ref{sec:implementation} summarizes the implementation.  Lastly the numerical results are given in section \ref{sec:numerical_tests} and \ref{sec:vortex_ring_collision}.

\section{Navier-Stokes LGF solution on a uniform grid} \label{sec:uniform}

\subsection{Discretization on unbounded uniform grid}\label{sec:uniform_discretization}
The Navier-Stokes LGF algorithm developed by \citet{liska2016fast} is briefly reviewed in this section.  This algorithm solves the incompressible viscous Navier-Stokes equations subject to the exact far-field boundary conditions. In a non-dimensional form, the equations are are given by

\begin{subequations}\label{eq:NS}
\begin{align}
\frac{\partial {\mathbf{u}}}{\partial t}+{\mathbf{u}} \cdot \nabla{\mathbf{u}}=-\nabla p+\frac{1}{\Rey} \nabla^{2}{\mathbf{u}}, \\
\quad \nabla \cdot{\mathbf{u}}=0,\\
{\mathbf{u}}(\vx, t) \rightarrow {\boldsymbol{0}}, \ \ p(\vx, t) \rightarrow p_\infty \quad \text {as } \quad |\mathbf{x}| \rightarrow \infty, \end{align}
\end{subequations}
where $\mathbf{u}$ is the velocity field, $p$ is the pressure, and $\Rey$ is the Reynolds number.

\begin{figure}[t]
    \centering
    \begin{overpic}[width=.9\textwidth]{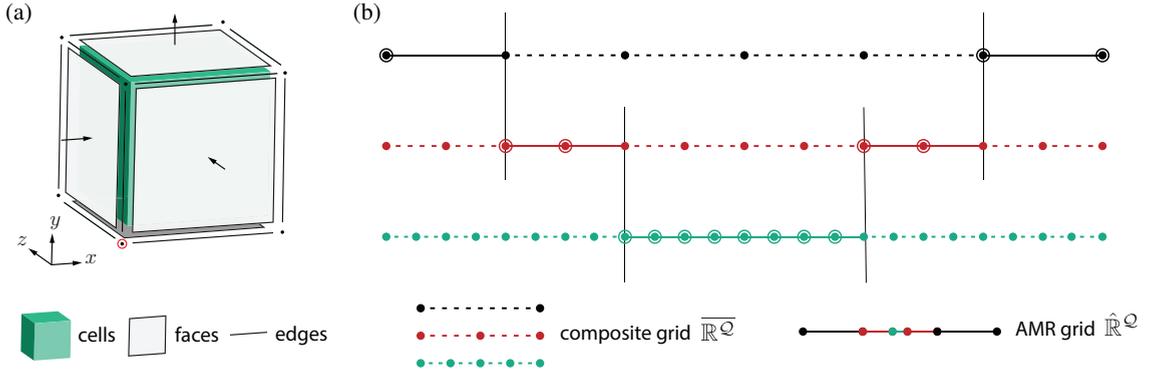}
        \put(-1,31) {\small (a)}
        \put (30,31) {\small (b)}
    \end{overpic}
    \caption{ (a) Base unit of a finite-volume staggered grid. (b) 1-D diagram for the composite grid (dotted line) and the corresponding AMR grid (solid line with corresponding vertices in circle)}\label{fig:grid}
\end{figure}

Eq.~\eqref{eq:NS} are formally discretized on an unbounded  staggered, uniform Cartesian grid (lattice) of single resolution. A base unit of this grid is shown in Fig.~\ref{fig:grid}a: its cell ($\mathcal{C}$) and vertices ($\cV$) discretize scalar quantities, and its positive faces ($\cF$) and edges ($\cE$) store vector quantities.  We use $\R^\cQ$ to denote the grid function spaces with values defined on $\cQ \in\{\mathcal{C}, \cF, \cE, \cV\}$.  The two principal discrete quantities to be solved for are the velocity and the pressure; we denote their corresponding grid functions (on the infinite lattice) as $\vu \in \R^{\cF}$ and $\vp \in \R^{\cC}$, respectively.

The grid function space is equipped with the following differential operators: the discrete gradient $\rmG:\R^{\mathcal{C}} \rightarrow \R^{\cF}$,
the discrete divergence $\rmD: \R^{\cF} \rightarrow \R^{\cC}$,
the discrete curl $\rmC: \R^{\cF} \rightarrow \R^{\cE}$ and $\overline{\rmC}: \R^{\cE} \rightarrow \R^{\cF}$, and the discrete Laplacian $\rmL_\cQ:\R^{\cQ} \rightarrow \R^{\cQ}$.
This discretization is of second-order accuracy and yields conservative, mimetic and commutative properties.
These properties are extensively exploited in this algorithm \cite{liska2016fast}. For instance, one has the following mimetic properties
\begin{align}
    \rmD = - \rmG^\dagger, \quad \overline{\rmC} = \rmC^\dagger, \quad \rmL_\cC = -\rmG^\dagger \rmG, \label{eq:mimetic}
\end{align}
and the following commutativity properties
\begin{align}
    \rmH_{\cF} \rmG = \rmG \rmH_{\cC},\quad   \rmD \rmH_{\cF} = \rmH_{\cC} \rmD, \label{eq:comm}
\end{align}
where $\rmH_{\cQ}$ is the integrating factor operator to be introduced in section \ref{sec:FIF-uniform}.

Using these differential operators, Eq.~\eqref{eq:NS} are discretized in space as
\begin{align}\label{eq:discrete-NS}
    \frac{\intdiff \vu}{\intdiff t}+\rmN\left(\vu+\vu_{\infty}\right)=-\rmG \vp+\frac{1}{\Rey} \rmL^{\cF} \vu, \quad \rmD \vu=0,
\end{align}
where $N\left(\vu+\vu_{\infty}\right)\approx \boldsymbol{\omega} \times\left(\mathbf{u}+\mathbf{u}_{\infty}\right)$ is the corresponding discretized non-linear term with $\vomega = \rmC \vu$ being the vorticity.  We use a second-order kinetic-energy preserving discretization of this term as reported in \cite{liska_fast_2017}.

%
%
%

\subsection{Fast LGF algorithm}\label{sec:FLGF-uniform}
Retaining the formally infinite grid, we employ the LGF for the corresponding discrete Poisson problem,
\begin{align}\label{eq:discrete_laplace}
&\rmL_\cQ  \vphi = \vf,  \quad  \lim_{n\rightarrow \infty} \vphi(\vn) = 0,\\
\Longrightarrow \quad &\vphi = \rmL_\cQ^{-1} \vf = G_\cQ * \vf,\label{eq:lgf_convolution}
\end{align}
where $\vf(\vn), \vs(\vn) \in \R^\cQ$, and $\vn$ denotes the trio of integers associated with the infinite lattice.  $G_\cQ$ is the LGF which incorporates exact far-field boundary condition and $*$ denotes the discrete convolution.

Given a source term $\vf(\vn)$ the solution can in principle be evaluated anywhere on the infinite lattice, but for a source with finite support, we only need do so at those lattice positions that are required to advance the solution.  This is accomplished in practice by thresholding the source of the Poisson equation, which is in turn proportional to the vorticity field.  Furthermore, this allows the solution to be spatially adaptive, as the active lattice points can be adjusted at each time-step.  This process is described in detail in \cite{liska_fast_2017}, and summarized later in section~\ref{sec:spatial_adaptivity}.

To accelerate the evaluation of Eq.~\eqref{eq:lgf_convolution}, a variant of the fast multipole method (FMM) is applied.  Specifically an FMM-based fast summation technique for a 3-D uniform Cartesian grid \citep{liska2014parallel} yields linear complexity and good parallel efficiency is employed.

\subsection{Integrating factor for the viscous term}\label{sec:FIF-uniform}

Similar to the LGF, an integrating factor (IF) is defined as the solution operator to the discrete heat equation on an unbounded uniform Cartesian grid
\begin{align}
    &\frac{\intdiff \vf}{\intdiff t}=\kappa \rmL_{\cQ} {\vf}, \quad {\vf}(\vn, \tau) = {\vf}_\tau(\vn), \label{eq:heat} \\
    \Longrightarrow\quad &{\vf}(\vn, t)=\left[\mathrm{\rmH}_{\cQ}\left(\frac{\kappa(t-\tau)}{(\Delta x)^{2}}\right) {\vf}_{\tau}\right](\vn, t), \quad t \geq \tau \label{eq:IF-single}
    ,
\end{align}
where $\vf\in \R^\cQ$, $\kappa >0$ is a constant,
$\vf_\tau(\vn)$ is a known source field, and $\rmH^{\cQ}$ is the integrating factor operator. The IF is a convolution with an exponentially decaying kernel, whereas the LGF kernel, $G_\cQ(\vn)$ in Eq.~\eqref{eq:lgf_convolution} decays as $1/|\vn|$.  The FLGF algorithm can be applied directly to this kernel \cite{liska2016fast}.

\subsection{Half-explicit Runge-Kutta scheme}\label{sec:runge-kutta_single}

With the IF technique permitting an exact time integration of the viscous term, the remaining terms are discretized in time using a half-explicit Runge-Kutta (HERK) method \citep{hairer2006numerical}.  HERK schemes exactly enforce algebraic constraints (in this case the divergence-free constraint), while using an explicit RK method to advance the differential equations.  More traditionally, split methods are needed so that the viscous terms are integrated implicitly, but the IF obviates this need.


By applying the IF operator $\rmH_{\cF}$ to Eq.~\eqref{eq:discrete-NS} we obtain
\begin{align}
    \frac{\intdiff \vv}{\intdiff t}=-\rmH_{\cF} \rmN\left((\rmH_{\cF})^{-1} \vv+\vu_{\infty}\right)-\rmH_{\cF} \rmG \vp, \quad \rmD (\rmH_{C})^{-1} \vv=0\label{eq:discrete-NS-IF},
\end{align}
where $\mathrm{v}=\rmH_{\cF} \vu \in  \R^\cF$. Because the integrating factor $\rmH_{\cF}$ commutes with the gradient and the divergence operators $\rmG $ and $ \rmD$ by Eq.~\eqref{eq:comm}, Eq.~\eqref{eq:discrete-NS-IF} is simplified to
\begin{align}\label{eq:discrete-NS-IF2}
    \frac{\intdiff \vv}{\intdiff t}=-\rmH_{\cF} \rmN\left(\rmH_{\cF}^{-1} \vv+\vu_{\infty}\right)-\rmG \rmH_{\cC} \vp , \quad \rmD \vv=0.
\end{align}

Eq.~\eqref{eq:discrete-NS-IF2} is integrated in time using the HERK scheme, which breaks down the total time integration into $N_k$ subproblems $\cup_{k=0}^{N_k} [t_k, t_{k+1}]$ . For each subproblem $[t_k, t_{k+1}]$, the HERK scheme takes an input of the velocity field $\vu(t_k)$, and output the velocity field $\vu(t_{k+1})$, which is defined as one timestep. The HERK scheme further breaks a timestep into many stages. At each stage it requires solving a system of equations of the following form
\begin{align}\label{eq:system-eqs}
    \left[\begin{array}{cc}{\left(\rmH_{\cF}^{i}\right)^{-1}} & {\rmG} \\ \rmD & {0}\end{array}\right]\left[\begin{array}{l}{\vu^{i}} \\ {\vp^{i}}\end{array}\right]=\left[\begin{array}{c}{\vr^{i}} \\ {0}\end{array}\right],
\end{align}
where $\vu=\rmH_{\cF}^{-1} \vv \in  \R^\cF$, $\vr$ is a known right-hand side, and the superscript refers to quantities evaluated at the $i$-th stage of the HERK scheme. Eq.~\eqref{eq:system-eqs} can be solved using a block-LU decomposition:
\begin{align}\label{eq:projection-long}
    \vu^{*}=\rmH_{\cF}^{i} \vr^{i}, \quad
    \rmD \rmH_{\cF}^{i} \rmG \vp^{i}= \rmD \vu^{*}, \quad
    \vu^{i}=\vu^{*}-\rmH_{\cF}^{i} \rmG \vp^{i} .
\end{align}
Again using the commutativity by Eq.~\eqref{eq:comm} and the mimetic properties by Eq.~\eqref{eq:mimetic}, Eq.~\eqref{eq:projection-long} is simplified to
\begin{align}
    \rmL_{\mathcal{C}}\vp^{i}&= \rmD \vr^{i},\label{eq:projection-1} \\
    \vu^{i}&=\rmH_{\cF}^{i}\left(\vr^{i}-\rmG \vp^{i}\right)\label{eq:projection-2}.
\end{align}
This simplified form involves the discrete Poisson equation, which is then solved with the FLGF technique discussed in section~\ref{sec:FLGF-uniform}.

On the right-hand side of Eq.~\eqref{eq:system-eqs}, $\vr^i$ is constructed using the information from the previous stages and the non-linear term at the current stage
\begin{align}
\vr^{i}&=\vq^{i}+\Delta t \sum_{j=1}^{i-1} \tilde{a}_{i, j} \vw^{i, j}+\vg^{i}, \label{eq:ri_recursive}
\end{align}
where $\Delta t$ is the time-step length, $\vg^{i}$ is related to the nonlinear term given by
\begin{align}
    \vg^{i}=-\tilde{a}_{i, i} \Delta t \, \rmN\left(\vu^{i-1}+\vu_{\infty}\left(t^{i-1}\right)\right), \quad  t^{i}=t+\tilde{c}_{i} \Delta t\label{eq:recursive-compute-2},
\end{align}
and $\vq^i$ and $\vw^{i,j}$ are recursively computed for $i > 1$ and $j < i$ using
\begin{align}
    &\vq^{i}=\rmH_{\cF}^{i-1} \vq^{i-1}, \quad \vq^{1}=\vu^{0} \\
    &\vw^{i, j}=\rmH_{\cF}^{i-1} \vw^{i-1, j}, \quad \vw^{i, i}=\left(\tilde{a}_{i, i} \Delta t\right)^{-1}\left(\vg^{i}-\rmG \vp^{i}\right)\label{eq:recursive-compute-3},
\end{align}
with $c_i$ and $\tilde{a}_{i, j}$ being the coefficients of a HERK scheme, and $\vu^{0}$ being the velocity field at the beginning of the time-marching.

The current implementation uses a HERK scheme introduced by \cite{brasey_half_explicit_1993} with the coefficients given in Table~\eqref{tab:HREK}, which corresponds to the Scheme B discussed in \cite{liska2016fast}. This scheme was chosen because it offers the highest order of accuracy for both the solution variable, velocity $\vu$ (third-order) and the constraint variable, pressure $\vp$ (second-order) among all three schemes considered in \cite{liska2016fast}.

\begin{table}[htbp]
    \centering
    \begin{tabular}{C|C C C}
    0 & 0 & 0 & 0 \\
    \frac{1}{3} & \frac{1}{3} & 0 & 0 \\
    1 & -1 & 2 & 0 \\
    \hline & 0 & \frac{3}{4} & \frac{1}{4}
    \end{tabular}
    \caption{Coefficients of the HERK scheme \cite{brasey_half_explicit_1993}.}
    \label{tab:HREK}
\end{table}

\section{Navier-Stokes LGF on an AMR grid}\label{sec:AMR}

As discussed in section~\ref{sec:uniform}, the mimetic properties of the differential operators and the commutativity between the differential operators and the LGF/IF operators are crucial for the algorithm (see the simplification of Eq.~\eqref{eq:discrete-NS-IF} and Eq.~\eqref{eq:projection-long}). Furthermore, the LGF is only defined on a regular grid and the regularity of the uniform grid also in turn enables an efficient evaluation of the fast LGF algorithm \cite{liska2014parallel}. However, an irregular grid does not possess those features. In this section
we propose a novel AMR technique that preserves the desired properties.

\subsection{Spatial discretization on an AMR grid}

In this section an AMR grid used for the discretization of Eq.~\eqref{eq:NS} is constructed in two steps.  First, we define a series of uniform unbounded staggered Cartesian grids. Each grid is of the form introduced in section~\ref{sec:uniform} but with different resolution $\{\R^\cQ_k\}_k$, where $k \in \mathbb{Z^+}$ refers to the resolution or grid {\it level}.  We use the convention that $\R^\cQ_0$ is the coarsest level and grid $\R^\cQ_{k+1}$ is generated by evenly dividing every grid unit $\R^\cQ_k$ into $2^d$ new units with $d$ being the dimension of Eq.~\eqref{eq:NS}, and denote $N_l$ as the maximum number of levels ($0 \leq k < N_l$).

We refer to the collection of uniform, unbounded grids as the {\it composite} grid, which is defined as a tensor product of the series of grids
\begin{align}
    \overline{\R^\cQ} \coloneqq \otimes_{k=0}^{N_l} \, \R^\cQ_{k}.
\end{align}
We equip this new tensor space  with an inner product that is induced from each $\R^\cQ_{k}$.

In the second step, an AMR grid is constructed as a subspace of the composite grid. More specifically, we define an AMR grid through a restriction operator. For each level $k$, a restriction operator $ \vGamma^{\cQ}_k \colon \R^\cQ_k \rightarrow  \R^\cQ_k$ is a linear functional defined as
\begin{align}
    \left(   \vGamma^{\cQ}_k\, \vf \right) (\vn)  =\left\{
    \begin{array}{cl}{
    \vf(\vn)} & {\text { for } \vn \in \Omega_k}  \\
    {0} & {\text { otherwise }}
    \end{array}\right.
\end{align}
for all $\vf \in \R^\cQ_k $, where  $\{\Omega_k\}_k$ are regions that partition the whole space $\R^3$, i.e.,
\begin{align}
    \Omega_l \cap \Omega_q = \delta_{lq} \Omega_l, \quad \cup_{q=0}^{N_l} \, \Omega_q = \R^3.\label{eq:partition}
\end{align}
For simplicity we write $\vGamma^{\cQ}_k$ as $\vGamma_k$ since $\cQ$ can be determined by the context.  A full restriction operator for the composite grid, $\vGamma \colon \overline{\R^{\cQ}} \rightarrow\overline{\R^{\cQ}}$ is then defined by the tensor product
\begin{align}
    \vGamma^\cQ \coloneqq \otimes_{k=0}^{N_l} \, \vGamma^\cQ_{k}.
\end{align}
We again simplify the notation $\vGamma^\cQ$  to be $\vGamma$. Finally an AMR grid $\hat{\R}^\cQ \subset \overline{\R^\cQ}$ is defined as the image of the full restriction operator $\vGamma$, i.e.,
\begin{align}
   \hat{\vf} \coloneqq \vGamma \,\vf \in \hat{\R}^\cQ, \quad \forall \vf \in\overline{\R^\cQ}.
\end{align}

\subsection{Interpolation/coarsening operators} \label{sec:P-operators}

So far we have defined a composite grid function space $\overline{\R^{\cQ}}$ which consists of $N_l$ unbounded uniform Cartesian grids, and an AMR grid $\hat{\R}^{\cQ}$ as a subspace of $\overline{\R^{\cQ}}$.  By definition, the full restriction operator projects functions in $\overline{\R^{\cQ}}$ to the AMR grid $\hat{\R}^{\cQ}$.  Numerically we only store information on the AMR grid.  Assuming every part of the AMR grid is approximating the same continuous function, then the information on the composite grid can be approximated using interpolations and coarsening. In this section we introduce the interpolation/coarsening operators that fulfill this idea.

First, we denote an interpolation/coarsening operator between any two levels $l$ and $q$ as $\vP^\cQ_{l\rightarrow q}$ ($\vP^\cQ_{l\rightarrow q}$ is an interpolation when $l<q$, it is coarsening when $l>q$, and it is the identity when $l=q$). We construct $\vP^\cQ_{l\rightarrow q}$ as compositions of the $\vP$-operators between consecutive levels.
For example, an interpolation operator between level $l$ and $q$ ($l<q$) is given by
\begin{align}\label{eq:P_composition}
    \vP^\cQ_{l\rightarrow q} =
    \vP^\cQ_{q-1 \rightarrow q}
    \cdots
    \vP^\cQ_{l+1 \rightarrow l+2}
    \vP^\cQ_{l \rightarrow l+1},
    \quad \text { for } 0\leq l<q< N_l.
\end{align}
This construction will be shown favorable for the numerical efficiency in section~\ref{sec:FLGF-AMR}.
Since the AMR grid is defined by the regions $\{\Omega_k\}_k$ that partition the space, then for a given level, the information can be estimated anywhere using the $\vP$-operators. More specifically, given $\hat{\vf}=\otimes_k \hat{\vf}_k \in \hat{\R}^\cQ \subset\overline{\R^\cQ}$ on the AMR grid, the information on level $k$ of the composite grid can be estimated by
\begin{align}
    \vf_k = \sum_{i=0}^{N_l} \vP^\cQ_{i\rightarrow k}  \hat{\vf}_i + O(h^{N_p}) \coloneqq \vP^\cQ_{k}\, \hat{\vf} + O(h^{N_p}), \label{eq:interp}
\end{align}
where the error convergence rate, $N_p$, is determined by the specific choice of interpolant. This is discussed in more detail below. In the second equality in Eq.~\eqref{eq:interp}, we defined another interpolation/coarsening operator  between the AMR grid and the uniform grid on level $k$, $\vP^\cQ_{k} : \hat{\R}^\cQ \rightarrow \R^\cQ_{k}$.  Similarly, a full $\vP$-operator between the AMR grid and the composite grid $\vP^\cQ\colon \hat{\R}^\cQ \rightarrow \overline{\R^\cQ}$ is defined by
\begin{align}
    \vP^\cQ = \otimes_{k=0}^{N_l} \, \vP^\cQ_{k}.
\end{align}
Together with the restriction operator $\vGamma$, one has the approximation relation between the AMR grid and the composite grid
\begin{align}
    \hat{\vf} &= \vGamma\, \vf, \quad \vf \approx \vP^\cQ \,\hat{\vf}, \label{eq:approx}
\end{align}
where $\vf \in \overline{\R^{\cQ}}$ and $\hat{\vf}\in \hat{\R}^{\cQ}$.

The general idea of our AMR technique is that, we consider the information on the AMR grid as being restricted from the ambient composite grid. At every time-step we can formally `recover' the information on the composite grid from the AMR grid using the $\vP$-operator. Then, the information on every level of the composite grid is formally marched in time even if the specific operations need not be performed on the entire composite grid. At the end of the time-step, the solution on the composite grid is again restricted to the AMR grid through the restriction $\vGamma$.  The most important feature of this process is that it need only be carried out on those portions of the composite grid that are needed to advance the AMR grid.  Thus, while in principle the solution is defined on every grid level, only the subspace defined by the AMR grid is required in practice.  We discuss how this is done in the next sections.

\subsection{Differential operators}\label{sec:diff_opr-composite}
Differential operators are simply constructed for the composite grid by
\begin{align}\label{eq:diff_operators}
    {\rmA} = \otimes_{k=0}^{N_l} \, \rmA_{k},
\end{align}
where $\mathrm{A_k}\in \{\rmG_k,\rmD_k,\rmC_k,\rmL_k\}$ are the corresponding discrete differential operators on grid $\R_k$ defined in section~\ref{sec:uniform_discretization}. This construction also preserves the second-order accuracy, conservation properties, mimetic properties, and commutativity (for the composite grid) since for every level they are the same as the native ones defined in section~\ref{sec:uniform_discretization}.  We note that differential operators need not be constructed directly for the AMR grid.

%
%
%
%

\subsection{Fast LGF alogirthm on the AMR grid}\label{sec:FLGF-AMR}

Before discussing the full algorithm for the Navier-Stokes equations, we provide details for applying the fast LGF/IF algorithms on the AMR grid by using the techniques derived in the proceeding section.  The resulting algorithm is essentially the same as aforementioned FLGF-AMR algorithm \cite{dorschner2020fast}. Here we use the composite-grid ansatz introduced above to reinterpret the algorithm and provide a more complete correction term than the one derived previously.

To solve the Poisson equation on the AMR grid, we use the information on the AMR grid to reconstruct the field on the composite grid, where the Poisson equation is hypothetically solved on every level through the FLGF technique, and the solution is then restricted back to the AMR grid.
However, computationally one only has access to the AMR grid. To efficiently evaluate the aforementioned process, we consider the commutativity between the interpolation and the LGF convolution: instead of interpolating the information from a coarse grid to a fine grid and then applying the LGF convolution, we seek to apply the LGF convolution on the coarse grid first and then interpolate the solution to the fine grid. In other words, we seek a commutative $\vP$-operator $\overline{\vP}^{\cQ}_{k\rightarrow l}$ that satisfies
\begin{align} \label{eq:comm_p}
\rmL_l^{-1}  \vP^{\cQ}_{k\rightarrow l} =
\overline{\vP}^{\cQ}_{k\rightarrow l}\rm\rmL_{k}^{-1},
\end{align}
where $0 \leq k,l<N_l$ denote to two distinct levels in the composite grid, and $\rmL_{k}$, $\rmL_{l}$ are the corresponding Laplacian operators.
Solving for $\overline{\vP}^{\cQ}_{k\rightarrow l} $ yields
\begin{align} \label{eq:p_bar}
    \overline{\vP}^{\cQ}_{k\rightarrow l}  = \rmL_{l}^{-1} \vP^{\cQ}_{k\rightarrow l} \rmL_k,
\end{align}
which suggests that $\overline{\vP}^{\cQ}_{k\rightarrow l}$ takes the form of a convolution.
Note that $\overline{\vP}^{\cQ}_{k\rightarrow l}$ holds a similar composition relation as Eq.~\eqref{eq:P_composition}
\begin{align}\label{eq:p_bar_comm}
    \overline{\vP}^{\cQ}_{k\rightarrow l} = \overline{\vP}^{\cQ}_{q\rightarrow l} \overline{\vP}^{\cQ}_{k\rightarrow q}.
\end{align}
A useful form of $\overline{\vP}^{\cQ}_{k\rightarrow l}$ from Eq. (\ref{eq:p_bar}) is derived by considering  $\overline{\vP}^{\cQ}_{k\rightarrow l}$ as the original interpolation $\vP^{\cQ}_{k\rightarrow l}$ with a correction
\begin{align}\label{eq:source_correct}
    \overline{\vP}^{\cQ}_{k\rightarrow l}  &=
    \rmL_{l}^{-1} \vP^{\cQ}_{k\rightarrow l} \rmL_k \nonumber \\
    &=\vP^{\cQ}_{k\rightarrow l} + \rmL_{l}^{-1} (\vP^{\cQ}_{k\rightarrow l} \rmL_k  - \rmL_{l} \vP^{\cQ}_{k\rightarrow l}) \nonumber \\
    & \coloneqq  \vP^{\cQ}_{k\rightarrow l} + \rmL_{l}^{-1} \vS^{\cQ}_{k\rightarrow l},
\end{align}
where the correction is in the form of a source term, given by applying the operator $\vS^{\cQ}_{k\rightarrow l}$ to the solution field on level $k$ (Eq.~\eqref{eq:comm_p}). One important property is that the correction source $\vS^{\cQ}_{k\rightarrow l} \rmL_{k}^{-1}$ yields a faster decay than the original LGF kernel.
For example, the correction term constructed for polynomial interpolations are shown to oscillate and decay as $|\vn|^{-4}$, whereas the LGF only decays as $|\vn|^{-1}$.\footnote{
As an example, the source correction constructed from an interpolation of simple averaging decays to about $3\times 10^{-5}$ at $20$ cells away from the center of the interpolation.}

\begin{figure}[htb]
    \begin{overpic}[width=.9\textwidth]{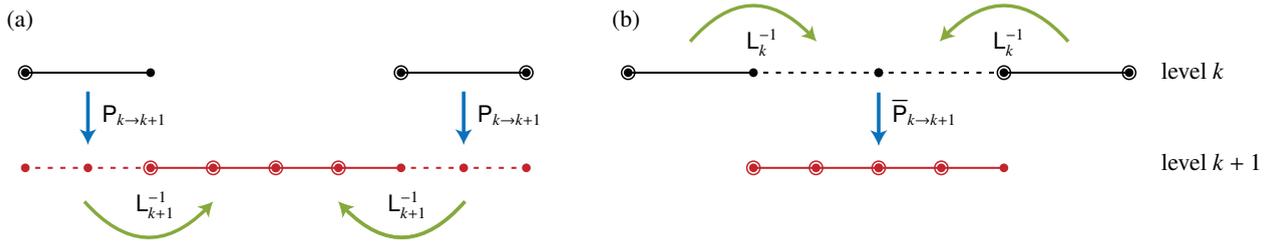}
        \put(-1,19) {\small (a)}
        \put (53,19) {\small (b)}
        \put (7.5,10.5) {\small $\vP_{k\rightarrow k+1}$}
        \put (41,10.5) {\small $\vP_{k\rightarrow k+1}$}

        \put (78,10.5) {\small $\overline{\vP}_{k\rightarrow k+1}$}

        \put (10.5,2.5) {\small $\rmL_{k+1}^{-1}$}
        \put (33,2.5) {\small $\rmL_{k+1}^{-1}$}

        \put (65,17) {\small $\rmL_{k}^{-1}$}
        \put (87,17) {\small $\rmL_{k}^{-1}$}

        \put (102,14.3) {\small level $k$}
        \put (102,6.2) {\small level $k+1$}
    \end{overpic}
    \caption{1-D diagram for the commutative interpolation with the Laplacian to avoid constructing a fine grid on level $k+1$: (a) applying an interpolation $\vP_{k\rightarrow k+1}$ first and $\rmL_{k+1}^{-1}$ second; (b) applying $\rmL_{k}^{-1}$  first and $\overline{\vP}_{k\rightarrow k+1}$ second.} \label{fig:commutativity}
\end{figure}

The Poisson equation for the composite grid $\overline{\R^{\cQ}}$ subject to the far-field boundary condition is given by
\begin{align}
    \rmL \vphi &= \vf, \quad \lim_{\mathbf{|n|} \rightarrow \infty} \vphi (\vn) = 0,
\end{align}
where $\vf, \vphi \in \overline{\R^{\cQ}}$ and $\rmL$ is the Laplacian for the composite grid defined in section~\ref{sec:diff_opr-composite}.
Approximating the source term on the whole composite grid $\vf$ using the AMR grid by Eq.~\eqref{eq:approx} and solving for $\vphi$ shows
\begin{align}
    \vphi(\vn) &= \rmL^{-1} \vf \nonumber \approx \rmL^{-1} \left( \vP^\cQ\,\hat{\vf}\right) \nonumber\\
    &= \otimes_{k=0}^{N_l} \, \rmL^{-1}_k \left(\sum_{i=0}^{N_l}  \vP^\cQ_{i\rightarrow k} \hat{\vf}_i \right)\nonumber \\
    &= \otimes_{k=0}^{N_l} \, \left[
    \left(\sum_{i=0}^{k-1}  \overline{\vP}^\cQ_{i\rightarrow k} \rmL^{-1}_i  \hat{\vf}_i \right) +
    \rmL^{-1}_k \left(\sum_{i=k}^{N_l}  \vP^\cQ_{i\rightarrow k} \hat{\vf}_i \right)
    \right] \nonumber \\
    &\coloneqq \otimes_{k=0}^{N_l} \, \left[ \vphi_1^k(\vn)+\vphi_2^k(\vn)  \right]\label{eq:AMR-LGF-2}\\
    \hat{\vphi}(\vn) &= \mathrm{\vGamma} \vphi(\vn)\label{eq:AMR-LGF}
\end{align}
where $\rmL^{-1} = \otimes_k \rmL_k^{-1}$ is the LGF for the composite grid, $\hat{\vf}\in \hat{\R}$ lives only on the AMR grid, $\vphi_1^k(\vn)$ is the partial solution corresponding to the source field from coarser levels, and $\vphi_2^k(\vn)$ is the partial solution corresponding to the source contribution from all finer levels as well as level $k$ itself.
Eq.~\eqref{eq:AMR-LGF-2} uses the commutative $\overline{\vP}^{\cQ}$ operator for the interpolations in $\vphi_1^k(\vn)$. As shown by the diagram Fig.~\ref{fig:commutativity}, the commutative construction avoids interpolating to a fine grid while yielding the same results.
Eq.~(\ref{eq:AMR-LGF-2}, \ref{eq:AMR-LGF}) are effectively the FLGF-AMR algorithm introduced in \cite{dorschner2020fast}.

To summarize, the revised FLGF-AMR algorithm is:
\begin{enumerate}
    \item From fine to coarse levels, evaluate the source term $\sum_{i=k}^{N_l}  \vP^\cQ_{i\rightarrow k} \hat{\vf}_i$ in $\vphi_2^k(\vn)$ through coarsening. Because $\vP$-operators are defined as compositions of consecutive levels (Eq.~\eqref{eq:P_composition}), this term is calculated cumulatively by
    \begin{align}
        \sum_{i=k+1}^{N_l}  \vP^\cQ_{i\rightarrow k} \hat{\vf}_i &=
        \vP^\cQ_{k+1\rightarrow k} \left(\sum_{i=k+2}^{N_l}  \vP^\cQ_{i\rightarrow k+1} \hat{\vf}_i  + \hat{\vf}_{k+1}\right).
    \end{align}

    \item $\vphi_1^k(\vn)$ are also evaluated cumulatively but from coarse to fine levels
    \begin{align}
        \vphi_1^k(\vn) &= \sum_{i=0}^{k-1}  \overline{\vP}^\cQ_{i\rightarrow k} \rmL^{-1}_i  \hat{\vf}_i \nonumber\\
        &=
        \overline{\vP}^\cQ_{k-1\rightarrow k}\left[\rmL_{k-1}^{-1} \hat{\vf}_{k-1} + \sum_{i=0}^{k-2} \overline{\vP}^\cQ_{i\rightarrow k-1} \rmL^{-1}_i \hat{\vf}_i \right] \nonumber \\
        &\coloneqq \overline{\vP}^\cQ_{k-1\rightarrow k} \psi_{k-1} \label{eq:partial_LGF}
    \end{align}
    Using the source correction introduced in Eq.~\eqref{eq:source_correct}, Eq.~\eqref{eq:partial_LGF} becomes
    \begin{align}
        \vphi_1^k(\vn) =\vP^\cQ_{k-1\rightarrow k} \psi_{k-1}  + \rmL^{-1}_k \vS^\cQ_{k-1\rightarrow k}\psi_{k-1}. \label{eq:Scorrection}
    \end{align}
    This form avoids specific evaluation of the non-local $\overline{\vP}$-operators, i.e., source terms can be combined with $\vphi_2^k(\vn)$ and Eq.~\eqref{eq:AMR-LGF-2} can be directly evaluated as
    \begin{align}
    \vphi(\vn)
    &= \otimes_{k=0}^{N_l} \, \left[
    {\vP}^\cQ_{k-1\rightarrow k} \psi_{k-1}  +
    \rmL^{-1}_k \left( \vS^\cQ_{k-1\rightarrow k} \psi_{k-1} + \sum_{i=k}^{N_l}  \vP^\cQ_{i\rightarrow k} \hat{\vf}_i \right)
    \right], \label{eq:combine_S}
\end{align}
    where $\psi_{k-1}$ is evaluated accumulatively from coarse to fine levels and $\rmL_k^{-1}$ is evaluated using the FLGF algorithm for the uniform grid. We define the combined source in Eq.~\eqref{eq:combine_S} as $\vS_k$ given by
    \begin{align}
        \vS_k
        &=  \vS^\cQ_{k-1\rightarrow k} \psi_{k-1} + \sum_{i=k}^{N_l}  \vP^\cQ_{i\rightarrow k} \hat{\vf}_i, \label{eq:combined_source}
    \end{align}
    which is utilized in the refinement indicator function to be introduced in section \ref{sec:refinement_adaptivity}.

    \item Lastly, the restriction operator $\vGamma$ is applied by limiting the region needed for the interpolation and the region used in the FLGF algorithm. With the construction of the composite grid, AMR grid, $\vP$-operators, the restriction operator $\vGamma$, as well as the differential operators, the FLGF-AMR algorithm in a nutshell is
\begin{align}
\left[ \vGamma\, \rmL^{-1} \vP^\cQ\right] \hat{\vf}\label{eq:FLGF-AMR} .
\end{align}
\end{enumerate}

\begin{figure}[tb]
    \centering
    \begin{overpic}[width=.55\textwidth]{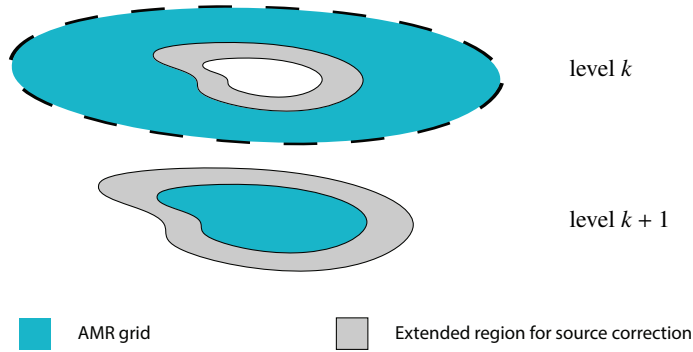}
        \put(82,40) {\small level $k$}
        \put (82,18) {\small level $k+1$}
    \end{overpic}
    \caption{ 2D Diagram for the AMR grid (blue) and the extended correction region (gray) at each refinement level.}\label{fig:source_correction}
\end{figure}

The approach developed here clarifies that the correction procedure developed in \cite{dorschner2020fast} is associated with the non-commutativity between $\vP$-operators and the LGF operators, and it yields a more precise form of the correction as an additive source term.
Though the source correction $\vS^\cQ_{k-1\rightarrow k}\psi_{k-1}$ decays rapidly, it is non-local and requires an extended region up to a cut-off distance depicted in Fig.~\ref{fig:source_correction}.  More specifically the AMR grid is defined with a set of physical domains $\{\Omega_k\}_k$ (Eq.~\eqref{eq:partition}), and the corresponding extended correction regions on level $k$ are given by
\begin{align}
\Omega_{k}^E=\left\{\vx:|\vx-\vy| \leq N_E \Delta x_k,\, \vy \in \Omega_{k}, \, \vx \notin \Omega_{k}\right\}, \label{eq:extended_regions}
\end{align}
where $\Delta x_k$ is the cell width on grid level $k$ and $N_E$ controls the cut-off distance of the extended region.

To test the new LGF-AMR algorithm, including the extended source correction, we use the same test case as \cite{dorschner2020fast} and solve a manufactured vorticity-streamfunction equation
\begin{align}
    \nabla^2 \Psi = \omega,
\end{align}
with the solution $\Psi$ given by
\begin{align}
\Psi(r, z)&=f\left(\frac{\sqrt{(r-R)^{2}+z^{2}}}{R}\right) \boldsymbol{e}_{\theta}, \qquad f(t)=\left\{\begin{array}{ll}
c_{1} \exp \left(-\frac{c_{2}}{1-t^{2}}\right) & \text { if } \quad|t|<1 \\
0 & \text { otherwise }
\end{array}\right.  .
\end{align}
For this test we let $c_1=10^3$, $c_2=10$ and $R=0.125$.  For every level, an extended correction region of a cut-off parameter $N_E=14$ is added, which corresponds to a relative source correction cut-off about $10^{-4}$.
We use the following criterion that a region on level $k$ is refined if
\begin{align}
    \omega^{k}(\vx) > \alpha^{L_{R}-k} \omega_{\max }, \quad \forall \vx \in \Omega_k,
\end{align}
where $L_R$ is the maximum number of refinement and  $\alpha=1/6$ is used.

\begin{figure}[ht]
    \centering
    \begin{overpic}[width=.95\textwidth]{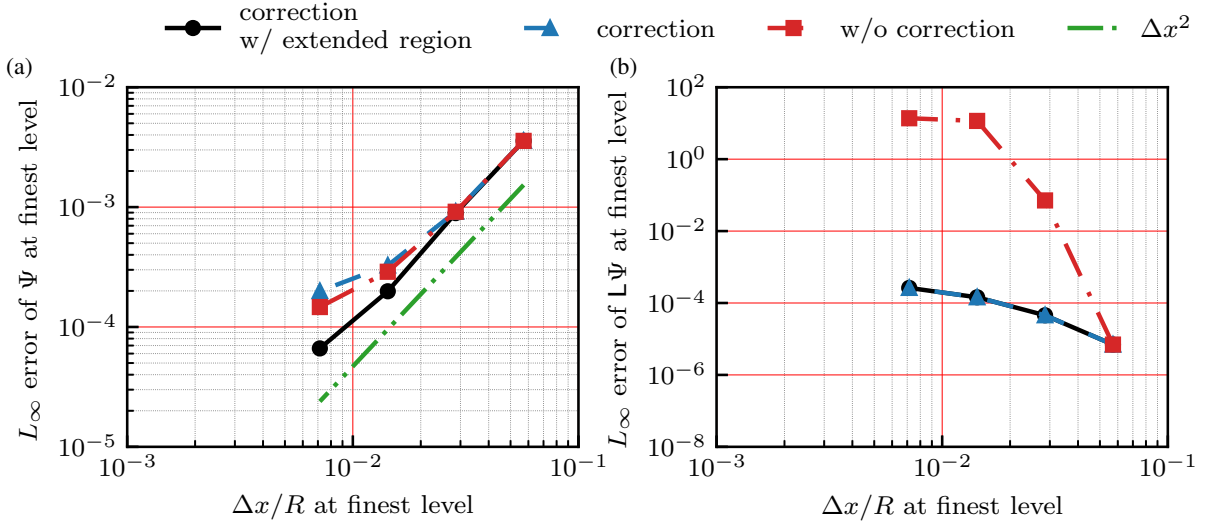}
        \put(-1,38) {\small (a)}
        \put(50,38) {\small (b)}
    \end{overpic}
    \caption{ Convergence of solutions on the finest grid w.r.t. to the refinement levels from $0$ to $3$  with a criterion $\alpha=1/6$ of (a) the numerical solution $\Psi$ and (b) left applying the discrete Laplacian to the numerical solution, $\rmL\Psi$. For each plot three cases are considered: without the correction (red); with correction but without an extended region (blue); and with correction and with an extended region (black). Across all three cases the mesh topology is kept the same. }\label{fig:source_correction_convergence}
\end{figure}

Fig.~\ref{fig:source_correction_convergence}a compares the $L_\infty$ error of the solutions on the finest grid level for an increasing number of refinement levels, and Fig.~\ref{fig:source_correction_convergence}b shows the error after left applying the discrete forward Laplacian $\rmL^Q$ to the numerical solutions.
Three cases are considered: (1) without source correction (2) with correction but without an extended correction region and (3) with correction and with an extended correction region.
For all tests the mesh topology is kept constant during the run.  It can be seen that the proposed extended source correction not only improves the accuracy of the solution but also helps ensure the consistency with the discrete forward Laplacian.
Note that on the every level, the extended source correction region is only added to the `source' in the FMM technique (section~\ref{sec:implementation}) but not the `target', so the asymptotic computation rate reported in \cite{dorschner2020fast} is not affected.

\subsection{Fast IF algorithm on the AMR grid}\label{sec:FIF-AMR}

Similar to the LGF for the Laplacian, the IF for the composite grid is constructed as $\rmH^{\cQ} = \otimes_k \rmH^{\cQ}_k$ with $\rmH^{\cQ}_k$ being the integrating factor for each level defined by Eq.~\eqref{eq:IF-single}. Similar to Eq.~\eqref{eq:FLGF-AMR} the fast IF-AMR (FIF-AMR) algorithm is
\begin{align}
    \hat{\vf}(\vn, t) \approx \left[ \vGamma\, \rmH^{\cQ}\, \vP^{\cQ}\right] \hat{\vf}_\tau(\vn),
\label{eq:FIF-AMR}
\end{align}
where $\hat{\vf},\hat{\vf}_\tau\in \hat{\R}^\cQ$ live on the AMR grid. Note that the kernel of $\rmH^{\cQ}$ decays exponentially, which simplifies the implementation as one only needs to apply the interpolation/coarsening $\vP$-operator to an extended region and then apply the IF convolution before finally restricting the solution back to the AMR grid. Numerically the extended region used for the FIF-AMR is the same as the one used for the FLGF-AMR shown in Fig.~\eqref{fig:source_correction}, and the same cut-off parameter $N_E$ for the extended region is used, which corresponds to a relative error less than $10^{-10}$ due to the exponential decay.

\subsection{Navier-Stokes FLGF-AMR-HERK algorithm }\label{sec:NS-LGF-AMR}

We now gather the elements developed in the preceding sections to construct the AMR technique for the full Navier-Stokes equations.  As discussed previously, the main steps are to (a) provide an algorithm to formally recover the flow field everywhere on the composite grid and time-marching every level, and (b) restrict the solution back to the AMR grid such that only a small subset of the composite grid need actually be computed.  To achieve this we combine the half-explicit Runge-Kutta scheme discussed in section~\ref{sec:runge-kutta_single} with the FLGF-AMR and FIF-AMR algorithms derived in section~\ref{sec:FLGF-AMR} and  section~\ref{sec:FIF-AMR}.

As we have constructed the differential operators and the LGF/IF operators on every level of the composite grid using the operators native to each unbounded uniform staggered grid, the mimetic properties and the commutativity are preserved. Thus Eq.~(\ref{eq:discrete-NS}-\ref{eq:projection-2}) are formally the same, but with the corresponding operators and variables referring to the those for the composite grid. For example, the LGF and IF operators for the composite grid is given by,
\begin{align}
    \rmL^{-1} = \otimes_{k=0}^{N_l} \, \rmL^{-1}_k, \qquad \rmH_{\cF}^{i}=\otimes_{k=0}^{N_l} \,
    \rmH_{\cF}\left(\frac{\left(\tilde{c}_{i}-\tilde{c}_{i-1}\right) \Delta t}{(\Delta x_k)^{2} \Rey }\right),\label{eq:composite-LGF-IF}
\end{align}
where the IF operator for the composite grid depends on the stage of the HERK scheme $i$, and the level of the composite grid $k$.
Like the FLGF-AMR algorithm described in section~\ref{sec:FLGF-AMR}, the NS-LGF-AMR-HERK algorithm tries to approximate the right-hand-sides of each update equation on the {\it composite grid}, and restrict the solution back to the AMR grid in a way such that the full composite grid is never built, but rather only the regions that are required by the AMR grid.

The process of evaluating the equivalent form of the system of equations Eq.~\eqref{eq:system-eqs} for the composite grid can be broken down as follows.
\begin{enumerate}
    \item Eq.~\eqref{eq:system-eqs} for the composite grid is also solved using the block-LU decomposition. Similarly because of the commutativity between the \textit{composite grid} differential operators and the IF operator, Eq.~(\ref{eq:projection-1}, \ref{eq:projection-2}) can formally be solved with the composite grid LGF and IF given by Eq.~\eqref{eq:composite-LGF-IF}. Again, in this framework we approximate only the right-hand sides using the information on the AMR grid
    \begin{align}
        \rmD \vr^i &= \vP^\cC\,\widehat{ \rmD r^i} + \vepsilon_\rmD, \label{eq:approx0} \\
        \vr^{i}-\rmG \vp^{i} &= \vP^\cF \left(\,\widehat{\vr^{i}}-\widehat{\rmG \vp^{i}}\right) +\vepsilon_\vu,  \label{eq:approx1}
    \end{align}
    where   $\widehat{\rmD r^i}=\vGamma \rmD r^i$ and  $\,\widehat{\vr^{i}}-\widehat{\rmG \vp^{i}}=\vGamma \left( \vr^{i}-\rmG \vp^{i} \right)$ are the restricted fields on the AMR grid, $\vP^\cC$ and  $\vP^\cF$ are the interpolation and coarsening operators, and the associated approximation error terms $\vepsilon_\rmD$ and $\vepsilon_\vu$ are
    \begin{align}
        \vepsilon_\rmD= \rmD \vr^i -\vP^\cC\,\widehat{ \rmD r^i}, \qquad \vepsilon_\vu= \left(\vr^{i}-\rmG \vp^{i}\right)  - \left[ \vP^\cF \left(\,\widehat{\vr^{i}}-\widehat{\rmG \vp^{i}}\right) \right].
    \end{align}
    These errors are controlled by the order of the interpolation/coarsening and the local grid resolution, which are discussed in more detail in the next section.

    \item Notice that the solution to the pressure Poisson equation is only used in the form $\widehat{\rmG \vp^{i}}$ and with Eq.~\eqref{eq:projection-1} the solution after the restriction is
    \begin{align}
        \widehat{\rmG \vp^{i}} = \vGamma \rmG \vp^{i} = \left[ \vGamma \rmG \rmL_{\mathcal{C}}^{-1}  \vP^\cC \right] \widehat{ \rmD r^i} + \widetilde{\vepsilon}_\rmD, \qquad \widetilde{\vepsilon}_\rmD = \vGamma \rmG\rmL_{\mathcal{C}}^{-1}\vepsilon_\rmD.
        \label{eq:approx3}
    \end{align}
    Here the FLGF-AMR algorithm given by Eq.~\eqref{eq:FLGF-AMR} is used. Note that because of the gradient operator $\rmG$ before the restriction, solutions from the FLGF-AMR are restricted to a grid which is one extra cell larger along the boundary on every level of the AMR grid.

    \item After solving the pressure gradient $\widehat{\rmG \vp^{i}}$, by Eq.~\eqref{eq:projection-2}, the updated velocity field at stage $i$ after being restricted back to the AMR grid can be expressed as
    \begin{align}
        \hat{\vu}^{i}&=\left[\Gamma \rmH_{\cF}^{i} \vP^\cF\right]\left(\,\widehat{\vr^{i}}-\widehat{\rmG \vp^{i}}\right) +\widetilde{\vepsilon}_\vu , \qquad \widetilde{\vepsilon}_\vu = \Gamma \rmH_{\cF}^{i} \vepsilon_\vu.
        \label{eq:approx4}
    \end{align}
    Here the FIF-AMR algorithm given by Eq.~\eqref{eq:FIF-AMR} is applied.
\end{enumerate}

The aforementioned process requires right-hand sides $\widehat{\vr^i}$ and $ \widehat{\rmD \vr^i}$. The evaluation of $\widehat{\rmD \vr^i}$ on the AMR grid is essentially the same as $\widehat{\vr^i}$, except that a restriction that is one extra cell larger along the boundary on every level of the AMR grid is used due to the divergence operator. $\vr^i$ is recursively defined using the solutions from previous stages  by Eq.~\eqref{eq:ri_recursive} and the process of updating $\widehat{\vr^i}$ using the idea of the composite grid is described as follows.
\begin{enumerate}
    \item  $ \vq^{i}$  and $\vw^{i,j}$ on the required portion of the composite grid are approximated using the AMR solutions from previous stages,
    \begin{align}
        \vq^{i}&=\rmH_{\cF}^{i-1} \vq^{i-1}
        =\rmH_{\cF}^{i-1}\left(\vP^\cF\hat{\vq}^{i-1} + \vepsilon_\vq\right) ,\label{eq:apprx5}\\
        \vw^{i, j}&=\rmH_{\cF}^{i-1} \vw^{i-1, j}
             = \rmH_{\cF}^{i-1}  \left( \vP^\cF\hat{\vw}^{i-1,j} + \vepsilon_\vw\right) .\label{eq:apprx6}
    \end{align}
    where $ \vepsilon_\vq$ and $\vepsilon_\vw $ are the interpolation/coarsening errors.
    The solutions on the AMR grid are given by
    \begin{align}
        \hat{\vq}^{i} &= \left[ \vGamma \rmH_{\cF}^{i-1}\vP^\cF \right] \hat{\vq}^{i-1} + \widetilde{\vepsilon_\vq},
        \qquad  \widetilde{\vepsilon_\vq} = \Gamma \rmH_{\cF}^{i} \vepsilon_\vq,\\
        \hat{\vw}^{i, j}  &= \left[ \vGamma \rmH_{\cF}^{i-1}\vP^\cF \right]  \hat{\vw}^{i-1, j} + \widetilde{\vepsilon_\vw},
        \qquad  \widetilde{\vepsilon_\vw} = \Gamma \rmH_{\cF}^{i} \vepsilon_\vw.
    \end{align}
    Here the FIF-AMR algorithm given by Eq.~\eqref{eq:FIF-AMR} is again used.
    \item Similarly the nonlinear term $\widehat{\vg^i}$ on the AMR grid is given by
    \begin{align}
        \widehat{\vg^{i}}
        &=
        -\tilde{a}_{i, i} \Delta t \vGamma\,\left[ \vP^\cF \, \mathrm{N}
        \left( \vu^{i-1}+\vu_{\infty}\left(t^{i-1}\right)\right) + \vepsilon_\rmN \right],
    \end{align}
    with
    \begin{align}
        \vepsilon_\rmN = \mathrm{N}\left(\vu^{i-1}+\vu_{\infty}\left(t^{i-1}\right)\right) -\vP^\cF \, \mathrm{N}
        \left( \hat{\vu}^{i-1}+\hat{\vu}_{\infty}\left(t^{i-1}\right)\right), \label{eq:approx7}
    \end{align}
    which contains two sources, an interpolation/coarsening error and an aliasing error.
\end{enumerate}

In summary, the NS equations are formally discretized on the composite grid and time integrated using the HERK scheme. At each stage of the RK time integration, the system of equations Eq.~\eqref{eq:system-eqs} for the composite gird is solved using the block-LU decomposition. By approximating the right-hand sides using the information from the AMR grid, the resulting algorithm applies the FLGF-AMR algorithm for the pressure gradient $ \widehat{\rmG \vp^{i}}$, and uses the FIF-AMR algorithm for the updated velocity $\hat{\vu}^{i}$, and the intermediate fields $\hat{\vq}^{i}$ and  $\hat{\vw}^{i, j}$.

\subsection{ Approximation errors } \label{sec:error}

The four interpolation/coarsening errors ${\vepsilon}_\vu$, ${\vepsilon}_\rmD$, ${\vepsilon}_\vq$ and ${\vepsilon}_\vw$ control the difference between the composite-grid and AMR-grid solutions.  For the composite grid, the method inherits the second-order convergence properties associated with the existing FLGF-HERK algorithm; these were characterized and measured in previous work \cite{liska2016fast}.
The additional errors associated with AMR have a local truncation error of order $O(\Delta x_k^2)$. This error will vanish subjected to global refinement of all levels together. We empirically demonstrate the convergence in section~\ref{sec:numerical_tests} by considering the evolution of a vortex ring.

Some additional observations can be made about the truncation errors.
By Eqs.~(\ref{eq:approx0}) to (\ref{eq:approx7}), the approximation errors can propagate from coarse to fine levels through the LGF and IF convolutions $\rmL^{-1}_\cC$ and $\rmH_\cF$ to produce
$\widetilde{\vepsilon}_\vu$, $\widetilde{\vepsilon}_\rmD$, $\widetilde{\vepsilon}_\vq$ and $\widetilde{\vepsilon}_\vw$. Meanwhile ${\vepsilon}_\rmN$ poses another source or the error due to non-linearity (aliasing).  As $\rmG \rmL^{-1}_\cC$ and $\rmH_\cF$ are bounded operators, the corresponding errors $\widetilde{\vepsilon}_\vu$ and $\widetilde{\vepsilon}_\rmD$ are well-behaved (the constant in the error term is finite).

Note that Eqs.~(\ref{eq:approx0}) and (\ref{eq:approx1}) simultaneously approximate the terms $\vr^i$ and $\rmD \vr^i$ using $\vP^\cF$ and $\vP^\cC$.
A more consistent approach would be to only approximate $\vr^i$ using interpolation/coarsening and evaluate $\rmD \vr^i$ accordingly.  
This would require a commutative $\vP$-operator with the divergence operator
\begin{align}
   \rmD \vP^\cF = \vP^\cC \rmD,
\end{align}
which could then be implemented using a similar correction step as in Eq.~\eqref{eq:Scorrection}. Unfortunately, this approach requires a construction of a divergence operator for the AMR grid---in our simplified approach operators need only be defined on the composite grid, and we have thus presently opted for the former, simpler approach (by tolerating the additional error term).

A final point regarding these errors is that they are, in principle, not different from those associated with any AMR scheme.  The spirit of AMR is to adapt the mesh according to the (measured) smoothness of the solution.  In other words, the AMR scheme will attempt to minimize all the errors discussed in this section, subject to being balanced by the overall truncation error.  This is achieved by the adaptation strategies described in the next section.




\section{Adaptivity}\label{sec:adaptivity}

We have thus far described the algorithm for solving the incompressible Navier-Stokes equations on an AMR grid by introducing approximations to the (theoretical) solution on a composite grid associated with interpolation/coarsening operators between grid levels, and appropriate restriction operators. The AMR grid involves a collection of regions $\{\Omega_k\}_k$, where $k$ denotes the level of refinement. The collection $\{\Omega_k\}_k$ still contains at least one unbounded region (i.e. the coarsest grid).
In this section we will discuss how to {\it adaptively} truncate the coarsest grid level and strategies for determining the adaptive truncation and adaptive refinement.

\subsection{Adaptive truncation }\label{sec:spatial_adaptivity}

The vortical regions in external flows are associated with the source term in either the pressure Poisson equation or the vorticity-streamfunction equation.
A truncation of the computation domain is plausible since
the vorticity field is compact, and it suffices to assume only the base level (coarsest grid) $\Omega_0$ is infinite.  The spatial truncation for the computational domain $\Omega_0$ is adapted from \citet{liska2016fast}, where a formally unbounded staggered uniform grid of a single resolution was also truncated. We refer to \cite{liska2016fast} for a more detailed explanation. Here we only provide a brief summary of the truncation algorithm and discuss how to combine it with the AMR technique.

Two types of convolutions are performed in the current algorithm, namely the LGF, $\rmL^{-1}_\cC$ for the pressure Poisson equation and the IF, $\rmH_\cF$ for the velocity field. To ensure the solution in the region of interest is both accurate and minimal in extent, the corresponding source terms in these convolutions need to be restricted to regions where they have magnitude greater than a tunable threshold value.

The source term in the LGF convolution introduced in section \S~\ref{sec:NS-LGF-AMR} is given by $\rmD \vr^i$. This term is approximately the divergence of the nonlinear term (Lamb vector), which is in turn proportional to the compact vorticity field. Given a threshold $\epsilon^*$, a truncation of the base level domain $\Omega^\mathrm{supp}_0$ for the source term $\rmD \vr$ is defined as
\begin{align}
    \Omega^\mathrm{supp}_0 \coloneqq \left\{\vx \in \R^3: \frac{|\rmD \vr(\vx)|}{\lVert \rmD \vr \rVert_\infty} \leq \epsilon^{*} \right\}. \label{eq:truncated_0}
\end{align}
$\Omega^\mathrm{supp}_0$ determines the domain needed to yield an accurate solution to the Poisson equation.

The source term in the IF convolution is the velocity field which however yields a much slower decay. To accurately and efficiently evaluate the IF convolution we make use of the following two properties: the kernel of the IF decays exponentially and the velocity field can be recovered from the vorticity. More specifically, to evaluate the solution of the IF convolution in the region of interest $\Omega_0^\mathrm{soln}$, only the velocity field in an extended region $\Omega_0^\mathrm{xsoln}$ is needed, which is defined by
\begin{align}
    \Omega_0^\mathrm{xsoln} \coloneqq \left\{\vx \in \R^3: \left| \vx-\vy \right| < d_{\mathrm{IF}}, \quad \vy \in \Omega_0^\mathrm{soln} \right\}, \label{eq:Xsoln}
\end{align}
where $d_{\mathrm{IF}}$ is a cut-off distance for the exponentially decaying kernel. The velocity in the extended region is recovered using the discrete vorticity-streamfunction relation
\begin{align}
\mathrm{u}=-\rmC^{\dagger} \rmL_{\cE}^{-1} \vomega   \label{eq:vel_update},
\end{align}
where $\vomega$ is the discrete vorticity, $\rmL_{\cE}$ is the Laplacian for the edge quantities, $\rmC^{\dagger}$ is the discrete curl for the edge quantities defined in section~\ref{sec:uniform_discretization}, and $\mathrm{u}$ is the discrete velocity.
We refer to this process as the velocity refresh. The velocity refresh is only needed for the base level and  the vorticity is calculated from the coarsened velocity field from the AMR grid given by
\begin{align}
     \vomega_0 = \rmC \vP_0^\cC \hat{u},
\end{align}
where $\hat{u}$ is the velocity on the AMR grid and $\vP_0$ is the coarsening operator defined by Eq.~\eqref{eq:interp}. The evaluation of Eq.~\eqref{eq:vel_update} uses the FLGF algorithm for a single level, introduced in section \ref{sec:FLGF-uniform}.
We also require $ \Omega_0^\mathrm{supp} \subset \Omega_0^\mathrm{soln} \subset \Omega_0^\mathrm{xsoln}$. The computational domain of the base level AMR grid $\Omega_0$ is truncated such that $\Omega_0^\mathrm{xsoln} = \cup_{k=0}^{N_l} \Omega_k$.

Lastly we note that the velocity refresh need not be performed at every time-step.  The frequency of the refresh depends on the decay of the IF kernel, and whether the base level mesh topology is updated. More details about the decay of IF kernel can be found in \cite{liska2016fast}.

\subsection{Adaptive refinement}\label{sec:refinement_adaptivity}

Adaptive refinement is achieved by updating the restriction operator to $\vGamma'$ as the solution progresses. The updated velocity field $\vu'$ on the new AMR grid is related to the original velocity field $\vu$ by interpolation/coarsening
\begin{align}
\vu'=\vGamma' \vP^{\cF} \vGamma \vu.
\end{align}
To yield an accurate solution, the resolution of the different parts of the AMR grid needs to reflect the different scales in the flow, which are a-priori unknown.

The choice of indicator function used to invoke refinement/derefinement has been discussed in previous work on AMR. For instance, \citet{berger1989} applies a Richardson extrapolation by comparing the time-marched solutions on both the coarse and fine mesh. The identification of vortical structures in the flow often relies on the usage of the gradient, the curvature and magnitude of the vorticity in the indicator functions \citep{almgren_conservative_1998, sussman_adaptive, popinet2003, sitaraman_parallel_2010}, while the gradient of the density is often used to detect the existence of shocks \citep{almgren_conservative_1998, quirk_parallel_1996, papoutsakis_efficient_2018}.
A combination of different indicators can also be used. For example, \citet{kamkar_automated_2011} uses the Q-criterion with the Richardson extrapolation and \citet{shenoy_unstructured_2014} uses the vorticity field, the non-linear term and the Q-criterion together.

Our NS-LGF-AMR-HERK scheme is mainly based on two algorithms: the FLGF-AMR algorithm for the pressure Poisson equation and the FIF-AMR algorithm for the velocity field involved in the viscous term. Both algorithms make use of the fundamental solutions defined on uniform grids 
by hypothetically interpolating/coarsening the source fields to the composite grid and solving on each level independently.  These interpolations/coarsening provide information about the truncation error and can therefore be used as an adaptation indicator.

The current implementation uses a refinement indicator function that focuses solely on the source term involved in the FLGF-AMR (Poisson) algorithm for the following two reasons. First the kernel of the IF yields an exponential decay, which results in a more localized error, whereas the LGF kernel decays much slower and the LGF convolution can instantly propagate the error to the whole flow field. Secondly the source term in the IF convolution, i.e., the velocity is smoother compared with the source term in the Poisson equation which is proportional to the vorticity field, and therefore it suffices to resolve the vorticity on the AMR grid.

More specifically, we propose to use the combined source term  $\vS_k$ from Eq.~\eqref{eq:combined_source} as the criterion.  Note that  $\vS_k$ is defined on the AMR grid as well as the extended source correction region introduced in section \ref{sec:FLGF-AMR}.  At time $t$, the AMR mesh at grid point $\vn$ on level $k$, or on level $k-1$ with an overlapping extended correction region on level $k$ is refined when
\begin{align}
   \vS_k(\vn, t) > \alpha^{N_l-k}  \vS_{\max} (t),  \label{eq:refine1}
\end{align}
where $0<\alpha<1$ is a constant, $N_l$ is the prescribed maximum refinement levels, and $\vS_{\max}(t)$ is a quantity that renders the criterion dimensionless. To make the refinement as efficient as possible, this quantity should monitor when the prescribed maximum resolution is most limited during the time horizon $[0, t]$ for an on-going simulation. The current implementation uses the following form
\begin{align}
   \vS_{\max} (t) = \max_{ \tau<t, \, B_n \in B} \mathrm{RMS}_{\vn \in B_n} \left[\vS_k(\vn, \tau)\right]  \label{eq:maximumS},
\end{align}
where $\mathrm{RMS}$ refers to the root mean square, $B_n$ denotes a block of computational cells, and $B$ denotes the union of cell blocks that partition the AMR grid. More details are discussed in section \ref{sec:implementation}. We choose the maximum rms of the combined source term over all cell blocks  because it estimates the least resolved region represented by the block where the maximum is reached, and as a statistical quantity it is less affected by numerical noise. Similarly, a region on level $k$ is coarsened when
\begin{align}
   \vS_k(\vn, t) < \beta \alpha^{N_l-k}  \vS_{\max} (t),  \label{eq:refine2}
\end{align}
where $0<\beta<1$ is a constant to avoid constant changes in the refinement levels due to small oscillations around the refinement criterion.

The combined source term $\vS_k$ consists of two parts:
the source term in the pressure Poisson equation $\rmD \vr^i$ defined in section~\ref{sec:NS-LGF-AMR}, and the source correction term defined in Eq.~\eqref{eq:source_correct}.
The former approximates the divergence of the non-linear term which is proportional to the vorticity field.
The later is related to the difference between the partial solutions (corresponding to the partial source from all coarser grids) on the coarse grid and the fine gird which is associated with the Richardson extrapolation. Using the combined source term has the benefits of an automatic incorporation of both the vorticity criterion and a Richardson extrapolation process using only one non-dimensionalization parameter without additional numerical expense.

\section{Implementation}\label{sec:implementation}


In this section the implementation of the Navier-Stokes LGF-AMR-HERK algorithm and the parallelization are briefly summarized. The development of the solver is based on \cite{dorschner2020fast} and the same data structure is adopted here. The solver is written in C++ and uses MPI for parallel communications.  The code uses a block-structured computational grid (i.e. the smallest unit for grid addition/removal and refinement/derefinement is a block of $N_b^3$ computational cells) and the blocks are also used in the refinement criterion Eq.~\eqref{eq:maximumS}.  The current implementation uses $N_b=N_E=14$. Those blocks are further organized using a tree structure (octree in 3D), where every node (octant) maps to a block.
The reason for using the octree is that the FLGF-AMR algorithm used in the NS-LGF-AMR-HERK scheme applies the FMM algorithm on each level of the computational grid, and each FMM operation uses the hierarchical subtree structure for the calculating the far-field interaction \cite{dorschner2020fast}. Different from \cite{dorschner2020fast}, where each leaf corresponds to a cubic domain in physical space, the current solver does not have this requirement since the extended correction regions can overlap with the physical space.

As in \cite{dorschner2020fast}, a server-client model is used for the  parallelization, where the server has the tree information but does not store any data, whereas each client only stores a part of the octree with its data.
At the beginning of a simulation, the server guesses a mesh topology according to the given initial condition, anticipates the load, and distributes the whole tree to the clients. While the work is mainly done by the clients during the run, the server receives the adaptivity requests (spatial addition/removal and local refinement/derefinement) from its clients, finds a new compatible mesh topology (smooth in the transition of refinement levels), calculates a new load distribution, and sends the adaptivity instructions back to the clients to transfer the data. The server is also responsible for identifying the subtree used in each FMM calculation during the FLGF-AMR algorithm.

The LGF kernel decays geometrically and is identical regardless of the grid level. Numerically, we store the exact values for near points, and use asymptotic expansions for points far away \cite{liska2014parallel, dorschner2020fast}. The IF kernel used in the AMR, on the other hand, depends on not only the stage of the HERK scheme, but also the grid level $k$ as suggested by Eq.~\eqref{eq:composite-LGF-IF}. However since it decays exponentially, one can still numerically calculate and store the exact values needed for all stages and levels during the initialization.

Since the IF kernel decays faster than the LGF, it requires fewer neighbor contributions.  Numerically, one evaluation of the FIF-AMR algorithm uses about $10\%$ of the execution time compared with the FLGF-AMR algorithm.
The three stage HERK scheme considered here requires the application of the FLGF-AMR algorithm at each stage, using around $60\%$ of the total execution time. The HERK scheme on the other hand applies the FIF-AMR algorithm to vector fields 5 times, contributing another $30\%$ of the total execution time.
One additional factor that affects the solver speed is the `velocity refresh' procedure introduced in section~\ref{sec:spatial_adaptivity} which requires solving the vector Poisson Eq.~\eqref{eq:vel_update} for the base level only. However the `velocity refresh' need not be performed at each time-step \cite{liska2016fast}. For the numerical tests considered in the following sections, we observe an additional contribution to the execution time of less than $15\%$.

\section{Verification} \label{sec:numerical_tests}

The Navier-Stokes AMR-LGF-HERK algorithm introduced in section~\ref{sec:NS-LGF-AMR} is verified by considering a fat-cored vortex ring with an initial vorticity distribution of the form
\begin{align}
    \omega_{\theta}(r, z)=\left\{\begin{array}{cl}
    \alpha \frac{\Gamma}{R^{2}} \exp \left(-4 s^{2} /\left(R^{2}-s^{2}\right)\right) & \text { if } s \leq R \\
    0 & \text { otherwise }
    \end{array}, \quad \omega_{z}(r, z)=0\right. ,
\end{align}
where $R$ is the radius of the vortex ring, $s^{2}=z^{2}+(r-R)^{2}$, and we let $\alpha \simeq 0.54857674$ so the total circulation integrates to the parameter $\Gamma$. The Reynolds number is set to $\Rey=\Gamma/\nu = 1000$, and the initial velocity field is calculated using the discrete vorticity-streamfunction relation Eq.~\eqref{eq:vel_update}.
The convergence study is performed by considering a series of runs with different levels of refinement and grid resolution: We vary the grid resolution on the base level from $\Delta x_{\text{base}}/ \Delta x_0 = 2^{-3},\cdots, 2^0$, where $\Delta x_0=R/4.7$ is a constant, and the grids are repeatedly refined using a refinement criterion $\alpha = 1/4$ (see section \ref{sec:implementation}) to reach the same finest resolution $\Delta x_{\text{fine}}/ \Delta x_0 = 2^{-3}$. The grid topology is kept constant during the test and all cases are performed with $\Delta t/ \Delta x_\text{base} = 0.35\times 2^{-N_l}$ up to $128$ time-steps for the finest case.
Finally the reference solution is performed using a uniform grid at double of the finest test resolution.

\begin{figure}[ht]
    \centering
    \begin{overpic}[width=.65\textwidth]{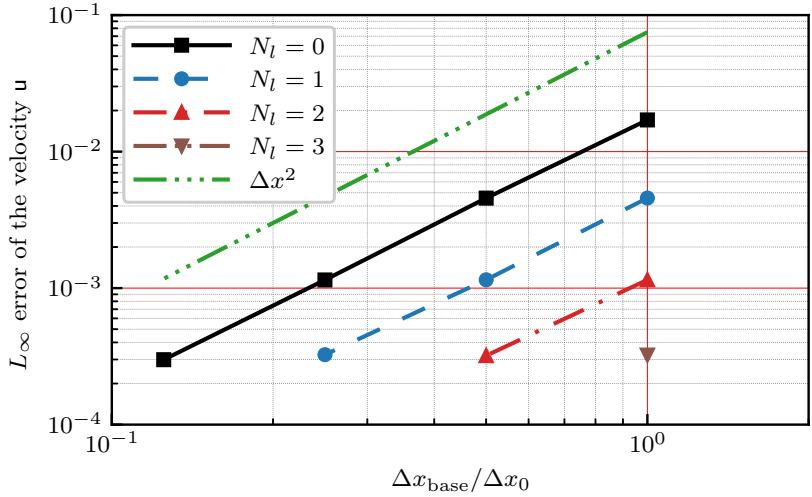}
    \end{overpic}
    \caption{ $L_\infty$ velocity convergence of the NS-FLGF-AMR solver wrt the grid resolution for refinement levels  $N_l=0,1,2,3$.}\label{fig:NS_convergence}
\end{figure}

The $L_\infty$ convergence in the velocity field for all ten cases are shown in Fig.~\ref{fig:NS_convergence}. The error is calculated by interpolating the reference solution onto the coarse grids using simple (2nd-order) averaging of nearest points. The plots show second-order convergence with the grid resolution for a fixed refinement level. Furthermore, with the aforementioned refinement criterion, adding a refinement level would produce a solution with comparable accuracy as refining the whole AMR grid at once. This verifies the efficacy of the AMR, i.e. we achieve better computational efficiency through local refinement. The computation saving in the spatial adaptivity and the nodal distribution will be further discussed in the section~\ref{sec:vortex_ring_collision}.

\section{Collision of vortex rings} \label{sec:vortex_ring_collision}

Vortex ring collisions are readily created in experiments \cite{oshima1978head, lim1992instability}. One notable feature of the vortex ring collision is that smaller secondary flow structures can develop at low Reynolds numbers, and turbulence cloud can form almost instantaneously at high Reynolds numbers, creating a wide range of length scales through a complex process of instabilities and vortex interaction \cite{mckeown2018cascade}.
The simulation of a vortex ring collision not only requires the numerics to be able to accurately and efficiently capture the fast and irregular changes in the flow configuration (e.g. the vortex ring radii can grow over 6 times during the expansion in the test case to be discussed), but at the same time also challenges the AMR scheme to add minimum noise to the flow as the transition is sensitive to perturbations.

In this section we use the collision of two thin vortex rings at high Reynolds number to demonstrate the Navier-Stokes FLGF-AMR-HERK solver with its spatial and refinement adaptivity.
We assume that each vortex ring has an initial vorticity profile
\begin{align}
    \omega_{\theta}(r, z)=\frac{\Gamma}{\pi \delta^{2}} \exp \left(-\frac{z^{2}+(r-R)^{2}}{\delta^{2}}\right), \quad \omega_{z}(r, z)=0,
\end{align}
where we set $\Rey=\Gamma/\nu=7500$, and $\delta/R=0.2$ controls the width of the vortex ring. The initial distance between the two vortex rings is set to $R+\delta$ to mitigate the initial interaction.
We consider three cases with the maximum refinement levels $N_l = 0, 1, 2$ respectively, and keep the resolution of the finest level the same with $\delta / \Delta x_{N_l} = 16$, with the ratio $\Delta t / \Delta x_{N_l} = 0.35$ held constant across all cases.
Both the spatial computational domain and the refinement regions are allowed to fully adapt. We use a spatial adaptive truncation threshold $\epsilon^*=10^{-4}$ defined in Eq.~\eqref{eq:truncated_0} and a refinement criterion $\alpha = 1/4$ and $\beta = 0.75$ defined in Eq.~(\ref{eq:refine1}, \ref{eq:refine2}).
Perturbations are added to the initial conditions of each vortex ring to accelerate the transition. The initial perturbation follows the recipe in \cite{shariff1994numerical}, where the radii of the two vortex rings are independently  perturbed with Fourier modes of uniform magnitude and random phases. For this test, perturbations are added to the first 32 modes with a magnitude of $3 \times 10^{-4}$ relative to the unperturbed vortex ring diameter $2R$.

\begin{figure}[ht]
    \centering
    \begin{overpic}[width=.9\textwidth]{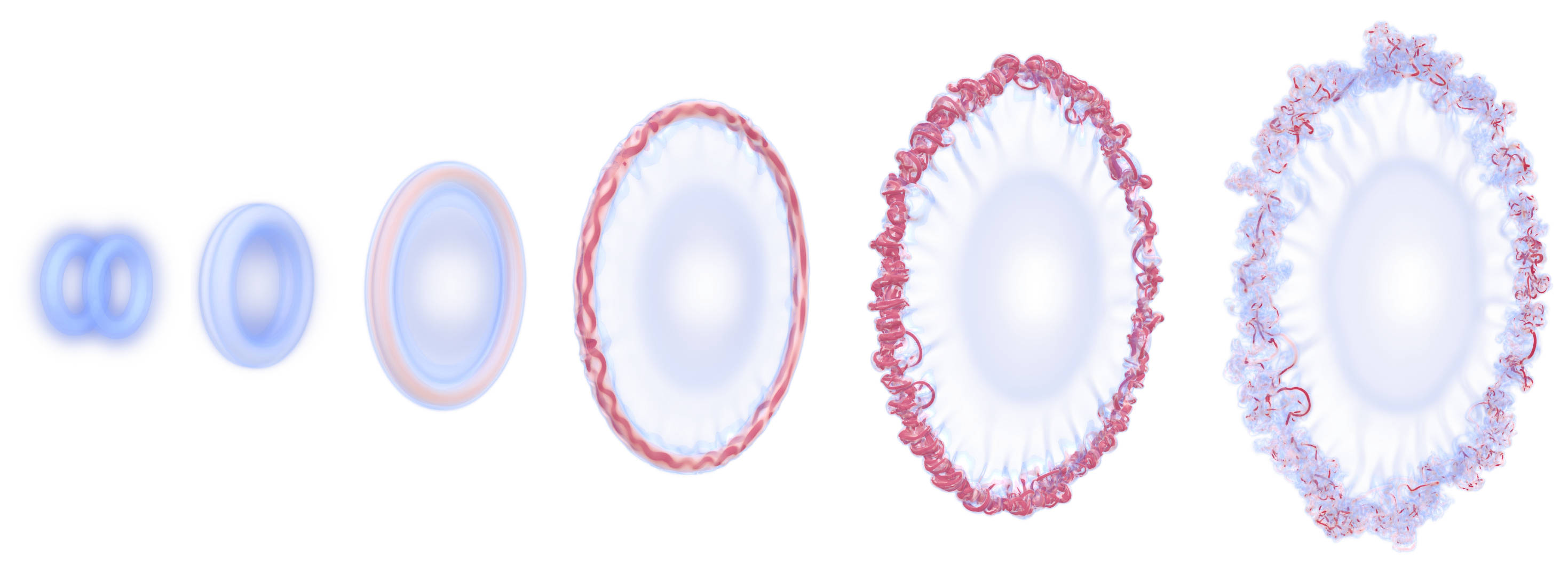}
    \end{overpic}
    \caption{
    Evolution of the vortex ring collision at $\Rey=7500$ from the case $N_l=1$. From left to right, vorticity isosurfaces $|\vomega R^2 / \Gamma |=[0.5-7.5]$ are given for $t \,\Gamma/ R^2 = $ 0, 4.5,  9.0,  15.1  19.9 and  25.2.}\label{fig:NS_collision}
\end{figure}

The evolution of the vortex ring collision taken from the case $N_l=1$ is shown in Fig.~\ref{fig:NS_collision}. 
As the rings collide, they expand rapidly about the impact center plane, leaving a  pair of thin vortex sheets trailing the leading vortices. The leading vortex pair becomes narrower, and as the vortex sheets are stretched, they eventually tear off, producing two disjoint circles. Both the Crow instability and the elliptic instability develop with the expansion which can be clearly observed around $t \,\Gamma/ R^2\sim 15$. The symmetry is broken, and finally the vortex ring pair transitions into turbulence.

\begin{figure}[tb]
    \centering
    \begin{overpic}[width=.9\textwidth]{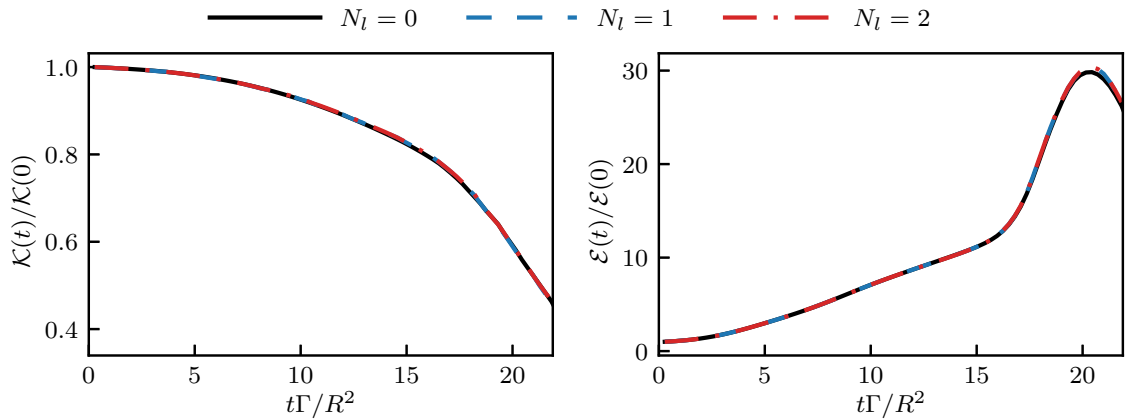}
    \end{overpic}
    \caption{ Evolution of the kinetic energy $\mathcal{K}(t)$ and enstrophy $\mathcal{E}(t)$ for the thin vortex ring collision at $\Rey=7500$ for maximum refinement levels $N_l=$0, 1 and 2. }\label{fig:stats}
\end{figure}

To quantitatively compare the AMR cases with the run $N_l=0$, we apply two statistical measures, namely the kinetic energy $\mathcal{K}(t)$ and the enstrophy $\mathcal{E}(t)$ given by
\begin{align}
    \mathcal{K}(t)&=\int_{\mathbb{R}^{3}} \mathbf{u} \cdot(\mathbf{x} \times \omega) d \mathbf{x}, \qquad \mathcal{E}(t)=\frac{1}{2} \int_{\mathbb{R}^{3}}|\boldsymbol{\omega}|^{2} d \mathbf{x}.
\end{align}
As shown in Fig.~\ref{fig:stats}, the expansion of the vortex ring pair is accompanied with a decay in the kinetic energy and a growth in the enstrophy. The turbulence transition starts around around $t\,\Gamma/ R^2\sim 15$ with an acceleration in enstrophy growth. The enstrophy reaches a maximum at $t\,\Gamma/ R^2\sim 20$, which corresponds to the fifth flow visualization in Fig~\ref{fig:NS_collision}. We see the results with varying numbers of refinement levels ($N_l=1, 2$) agree well with the uniform grid simulation ($N_l=0$) and predict the transition time accurately. As transitional flows are very sensitive to the noise, it suggests the extra numerical perturbation from the AMR scheme is lower than the initial perturbation.

\begin{figure}[ht]
    \centering
    \begin{overpic}[width=.99\textwidth]{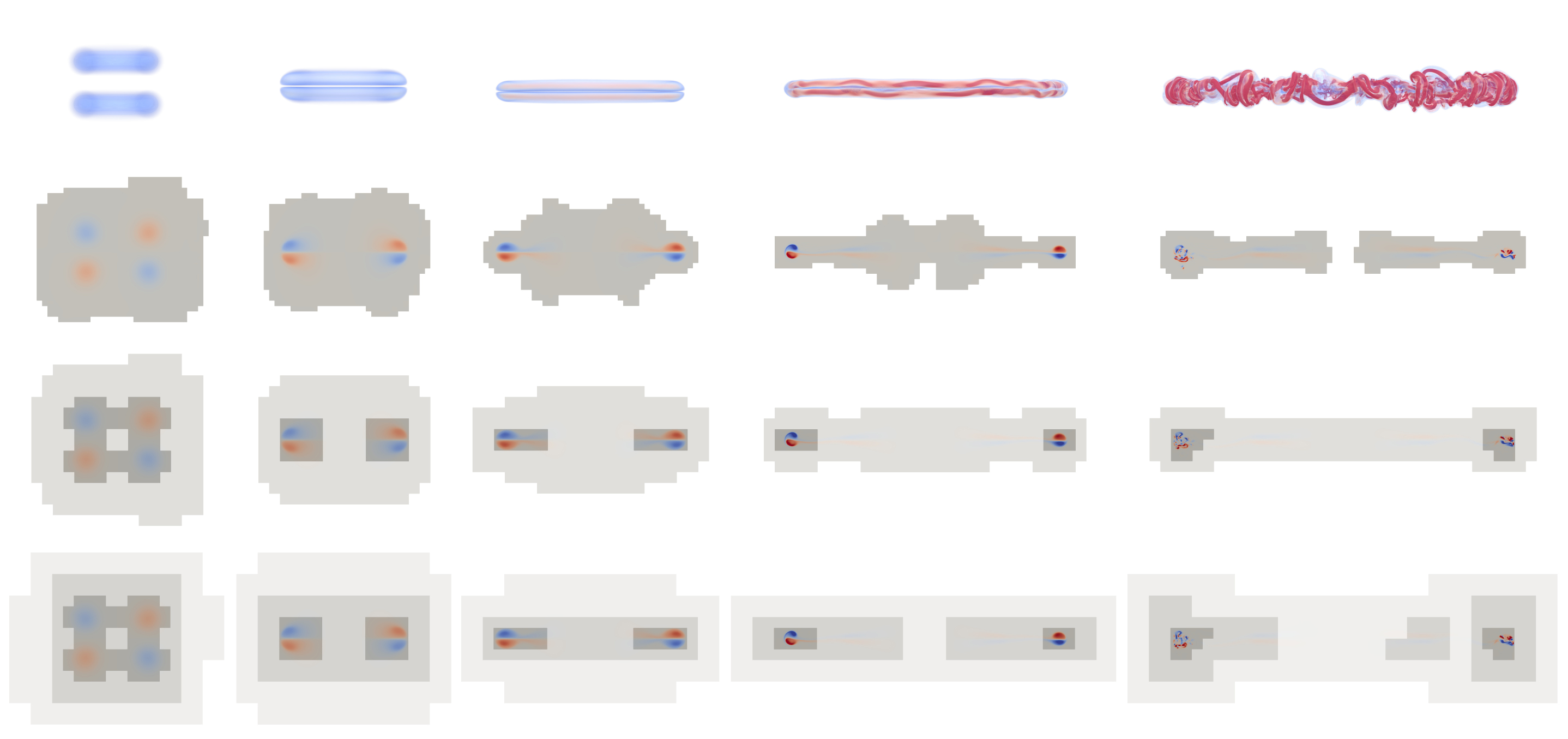}
    \end{overpic}
    \caption{ Mesh topology over the cross-section for $N_l=$0, 1 and 2. The top row shows the flow evolution at $|\vomega R^2 / \Gamma |=[0.5-7.5]$ are given for $t \,\Gamma/ R^2 = $ 0, 4.5,  9.0,  15.1 and 19.9. The second to the fourth row show the corresponding mesh topology for $N_l=0, 1, 2$. For all cases, coarse mesh to fine mesh are shown from light gray to dark gray.}\label{fig:meshTopologyLaminar}
\end{figure}

\begin{figure}[ht]
    \centering
    \begin{overpic}[width=.85\textwidth]{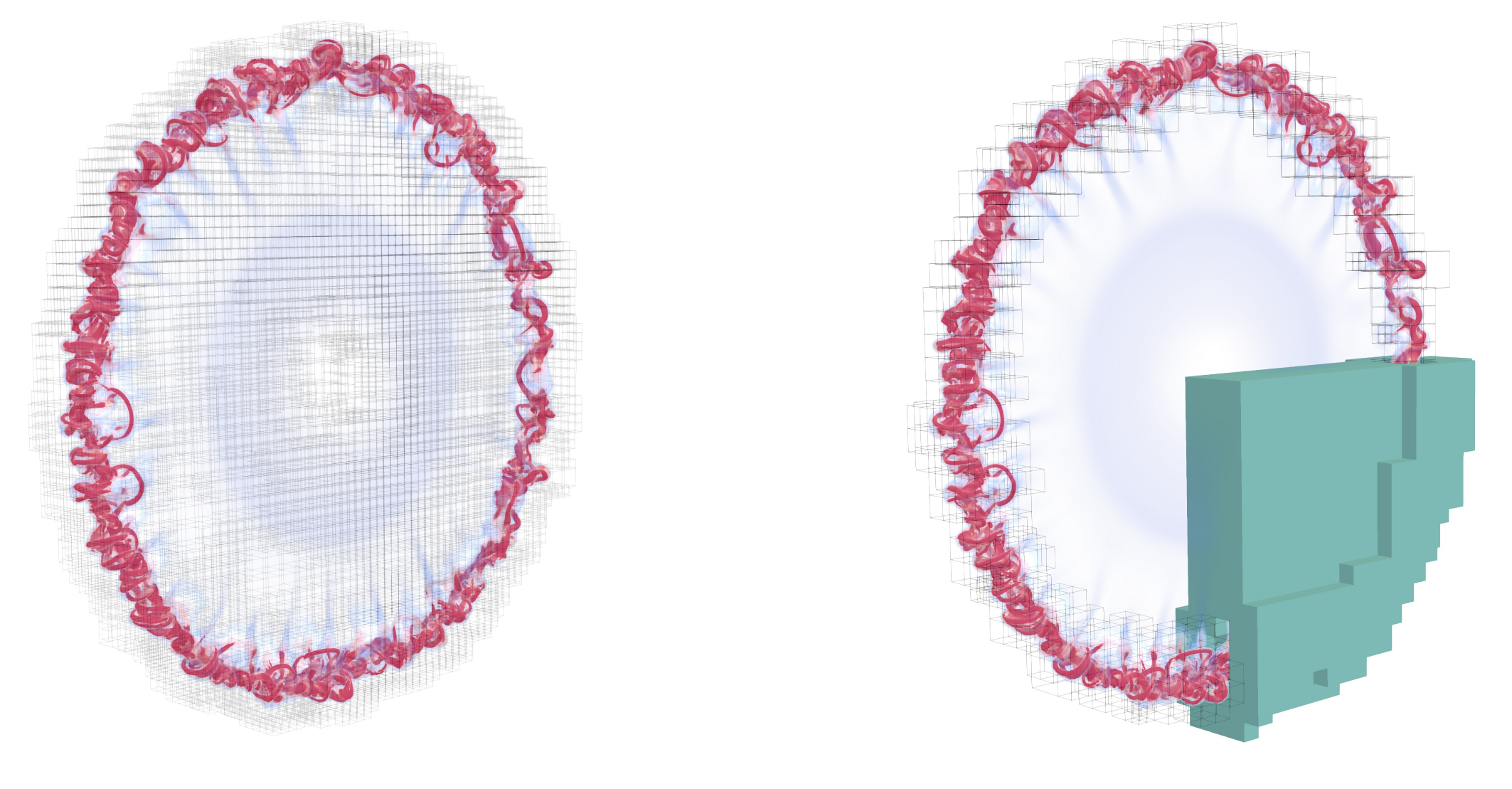}
        \put(-3,50) {\small (a)}
        \put(55,50) {\small (b)}
    \end{overpic}
    \caption{
     Flow visualization and mesh topology at $t \,\Gamma/ R^2=19.9$ for (a) the non-AMR case ($N_l=0$), and (b) the AMR case ($N_l=1$).  The computational blocks for the finest grid are shown in transparent boxes and a quarter of coarse grid for $N_l=1$ is shown in green.
    }\label{fig:mesh3D}
\end{figure}

Fig.~\ref{fig:meshTopologyLaminar} compares the mesh topology across the three cases for the time period ($t \,\Gamma/ R^2 = [0 - 20]$) where
the top row shows the side view of the same vorticity iso-surfaces as in Fig.~\ref{fig:NS_collision}, and  the second to the fourth rows show the mesh topology over a horizontal cross-section about the center. 
The flow fields and the mesh topology in 3D for the non-AMR case ($N_l=0$) and the AMR simulation ($N_l=1$) at $t \,\Gamma/ R^2 = 19.9$ are shown in Fig.~\ref{fig:mesh3D}, where we see the evolution compares well even after the entire transition period, and under the proposed criterion the refinement regions accurately surround the leading vortices while a lower resolution grid is used for the remaining regions.

The computational saving for all three cases is reported in Table~\ref{tab:saving}, where the numbers of cells used in the NS-FLGF-AMR-HERK scheme over the same time horizon are compared with a static rectangular domain of a minimum bounding box with the same finest resolution everywhere.  It can be seen that by involving the spatial adaptivity with the AMR, under the proposed criterion a factor of $10 \sim 20$ of reduction in computational cells is achieved.
\begin{table}[ht]
\centering
\begin{tabular}{L|LLLLL}
 t \,\Gamma/ R^2 & 0   & 4.5   & 9.0 & 15.1 & 19.9 \\
 \hline
N_l = 0 & 13.0\%  &12.9\% & 13.9\% & 17.0\%   & 24.0\%  \\
N_l = 1 & 5.2\%  &  4.7\% & 5.8\%   & 6.9\%  & 9.9\%  \\
N_l = 2 & 4.3\% & 4.5\%  & 4.5\%   & 5.4\%  & 9.7\% \\
\end{tabular}
\caption{ Number of computational cells for $N_l$=0, 1 and 2, compared with a static rectangular domain of minimum bounding box over the time horizon $t \,\Gamma/ R^2 = [0-20]$.}\label{tab:saving}
\end{table}

\section{Conclusions}

We proposed an AMR technique to enhance the LGF approach for solving viscous, incompressible flows on unbounded domains.
We consider the AMR grid as a subset of a composite grid that is constructed from a series of unbounded uniform staggered Cartesian grid of differing resolution.
Differential and LGF/IF operators are constructed for the composite grid and preserve the mimetic properties and the commutativity of the original IBLGF scheme. Interpolation/coarsening $\vP$-operators are defined through composition and their commutation with the aforementioned operators was studied. Based on this analysis, we refined the original FLGF-AMR algorithm \citep{dorschner2020fast} for solving the 3-D Poisson equation subject to  free-space boundary conditions.  The $\vP$-operators are applied to formally recover the information on the whole composite grid from the AMR grid, where the Poisson equation is hypothetically solved on every level using the FLGF technique, before the solution is restricted back to the AMR grid. We also showed that this hypothetical process can be evaluated efficiently by commuting the interpolation and the LGF convolution, resulting in an extended source correction and improved accuracy of the FLGF-AMR technique.

The Navier-Stokes equations were then discretized on this composite grid using a second-order FV scheme, and
we extended the AMR technique to incorporate the IF technique for the viscous term, and a Runge-Kutta (HERK) scheme for the resulting differential-algebraic equations. An incompressible Navier-Stokes method is then designed where we formally advance the flow fields on all levels of the composite grid using information from the AMR grid, but restricting the resulting computation back to the AMR grid obviates the need for all but a small subset of the composite grid.

Because the LGF represents the solution to the Poisson equation as a convolution of its source term, i.e. the divergence of the Lamb vector in the momentum equation, we construct an efficient and accurate refinement criterion that naturally tracks the associated truncation errors that are associated with interpolation and coarsening of the sources on different grid levels.

The Navier-Stokes solver was verified to give second-order accuracy through a refinement study of a fat-cored vortex ring.  We also demonstrated the capabilities and performance by simulating the collision of thin-cored vortex rings at $\Rey=7500$. We showed the AMR simulations agree well with the simulation using a uniform grid for the entire laminar and transitional period, while providing significant reductions in computational cells.

Lastly we note that the computational saving from the AMR depends on the scale separation present in the physics. In this work we restrict our attention to flows in the free space without bluff bodies. However as mentioned in the introduction, our AMR framework can readily be combined with the immersed boundary method, and a much higher reduction in computational expenses is expected as the AMR allows the thin boundary layers to be resolved more efficiently. This will be subject of future work.

\section*{Acknowledgments}
This work was supported by the ONR grant No. N00014-16-1-2734, the AFOSR/UCLA grant No. FA9550-18-1-0440 and the SNF Grant No. P2EZP2\_178436 (B. D.). This work used the Extreme Science and Engineering Discovery Environment \citep{towns2014xsede}, which is supported by National Science Foundation grant number ACI-1548562. Specifically, the computations presented here used Comet at the San Diego Supercomputer Center and Stampede 2 at the Texas Advanced Computing Center through allocation TG-CTS120005.


\newpage
\clearpage
\bibliographystyle{model1-num-names}
\bibliography{refs}

\end{document}